\documentclass[aps,prx,twocolumn,superscriptaddress,groupedaddress]{revtex4}
\usepackage[compat=1.0.0]{tikz-feynman}
\usepackage{amssymb}
\usepackage{multirow}
\usepackage{bm}
\usepackage{amsmath}
\usepackage{graphicx}
\usepackage{epstopdf}
\usepackage{subfigure}
\usepackage{natbib}
\usepackage{epsfig}
\usepackage{amsfonts}
\usepackage{mathrsfs}
\usepackage[toc,page,title,titletoc,header]{appendix}
\usepackage[colorlinks,linkcolor=blue,citecolor=blue,anchorcolor=blue]{hyperref}
\usepackage{dsfont,amsthm,amsbsy}
\usepackage[normalem]{ulem}

\def\Re{{\rm Re}}

\newcommand{\be}[1]{ \begin{eqnarray} \mbox{$\label{#1}$} }
\newcommand{\ee}{\end{eqnarray}}

\def\bea{\begin{eqnarray}}      \def\eea{\end{eqnarray}}
\def\ba{\begin{array} }
\def\ea{\end{array} }
\def\bnum{\begin{enumerate} }
\def\enum{\end{enumerate}}

\def\=>{\Rightarrow}
\def\>{\rightarrow}

\def\eye2{Fathbb{I}}

\renewcommand\d{\mathrm{d}}

\newcommand{\av} [1]{\langle #1 \rangle}
\newcommand{\eg}{{\it e.g. }}
\newcommand{\ie}{{\it i.e. }}

\begin{document}

\title{Two-Fluid Theory of Composite Bosons and Fermions \\
and the Quantum Hall Proximity Effect}
\author{Zhaoyu Han}
\affiliation{Department of Physics, Stanford University, Stanford, CA 94305, USA}
\author{Kyung-Su Kim} 
\affiliation{Department of Physics, Stanford University, Stanford, CA 94305, USA}
\author{Steven~A.~Kivelson}
\affiliation{Department of Physics, Stanford University, Stanford, CA 94305, USA}
\affiliation{Rudolf Peierls Centre for Theoretical Physics, University of Oxford, 1 Keble Road, Oxford OX1 3NP, United Kingdom}
\author{Thors Hans Hansson}
\affiliation{AlbaNova University Center, Department of Physics, Stockholm University, SE-106 91 Stockholm, Sweden}

\begin{abstract}
We propose a two-fluid description of fractional quantum Hall systems, in which one component is a condensate of composite bosons and the other a Fermi liquid formed by composite fermions (or simply electrons). We employ the theory to model the interface between a fractional quantum Hall liquid and a (composite) Fermi liquid metal, where we find a penetration of quantum Hall condensate into the metallic region reminiscent of the proximity effect in superconductor-metal interfaces. We also find a novel and physically reasonable set of gapped quasielectron and neutral modes in fractional quantum Hall liquids. 
\end{abstract}

\maketitle

\section{Introduction}

The qualitative physics of a $\nu=1/q$ (with $q$ odd) fractional quantum Hall (FQH) fluid is most simply understood from the analogy with superconductivity provided by the Chern-Simons-Ginzburg-Landau (CSGL) description of composite bosons (CB)~\cite{PhysRevLett.62.82,zhang1992chern}, while dissipative states - either an ordinary metal or a composite fermion (CF) liquid metal~\cite{PhysRevLett.63.199, PhysRevB.46.9889, PhysRevB.47.7312,PhysRevB.44.5246,jain2007composite} - require a fermionic description.  An investigation of the interface between the two --- and the extent to which a quantum Hall (QH) ``condensate'' can penetrate into a metallic region --- motivates us to develop a two-fluid description of a two-dimensional electron gas in a spatially varying magnetic field. To this purpose, we introduce an effective field theory with two fictitious flavors of electron - one described by a composite boson field and the other by a composite fermion field - coupled to two fluctuating Chern-Simons (CS) gauge fields.

There is an intrinsic issue with any such two-fluid descriptions stemming from the indistinguishability of electrons.  However, under many circumstances where the dynamical exchange of different groups of electrons is slow, errors involved in treating them as distinguishable are expected to be small. In the present treatment, the exchange statistics of each flavor of electrons is treated exactly through the statistical CS field.  Dynamical terms that exchange flavors involve instanton configurations in which large-scale rearrangements of the gauge fields in spacetime arise, which are argued to be relatively unimportant in various situations considered here. At an intuitive level, this is similar to treatments in which the electrons in a full Landau level (LL) are treated as forming an incompressible fluid background, over which electrons in a higher, partially filled Landau level form a distinct quantum system.

When the system is strictly uniform, and at special fillings that are well described by one component, the other component of the two fluids is clearly redundant. However, novel QH states can arise when a CB condensate and a CF liquid (CFL) coexist. Moreover, when the electron density or the magnetic field strength varies spatially, our mixed theory becomes useful in many circumstances. In particular, we apply it to two specific problems:

We first use the two-component approach to study point-like excitations of Laughlin states. While quasi-holes are plausibly describable as simple vortex excitations in the composite boson condensate, a correspondingly compelling CSGL description of quasielectrons is lacking~\cite{RevModPhys.89.025005}. Based on a saddle-point analysis, we construct a class of novel soliton solutions corresponding to quasielectrons and neutral excitations (a version of a Girvin-MacDonald-Platzman mode~\cite{PhysRevB.33.2481}), which do not admit a natural representation in terms of a pure CB picture. We find that depending on the details of the interactions, properties of the quasi-particles (\eg energy and quadrupole moment) can change discontinuously without any other significant changes in the nature of the QH state itself.

In the second application, we consider the interface between a $\nu=1/q$ (odd $q$) QH region and a metallic regime with $B=0$ or with $\nu=1/q'$ (even $q'$). Far from the interface, the two regions are well described by a CB condensate and a CFL (which is simply an electron liquid when $q'=0$), respectively.  For simplicity, we consider a situation in which the electron density is uniform and solve for the saddle-point profile of the CB condensate.   When the magnetic field is slowly varying, we find that the condensate roughly follows it, allowing it to extend deep into the predominantly metallic region, which is reminiscent of the proximity effect in superconductor-metal interfaces.

In Sec.~\ref{sec: formalism} we discuss the phenomenological considerations and the formalism of the proposed two-component effective field theory.  In Sec.~\ref{sec: uniform}, we briefly discuss the filling fractions that can potentially be described by the two-fluid picture. In Sec.~\ref{sec: quasiparticles}, we apply our formulation to point-like excitations in Laughlin states.  In Sec.~\ref{sec: interface}, we analyze the saddle-point solution of a problem with a spatially varying magnetic field modeling a QH-metal interface. Finally, in Sec.~\ref{sec: discussion} we speculate on the possible implications of these results and further extensions.

\section{Formalism}\label{sec: formalism}

\begin{figure}
    \centering
    \includegraphics[width = \linewidth]{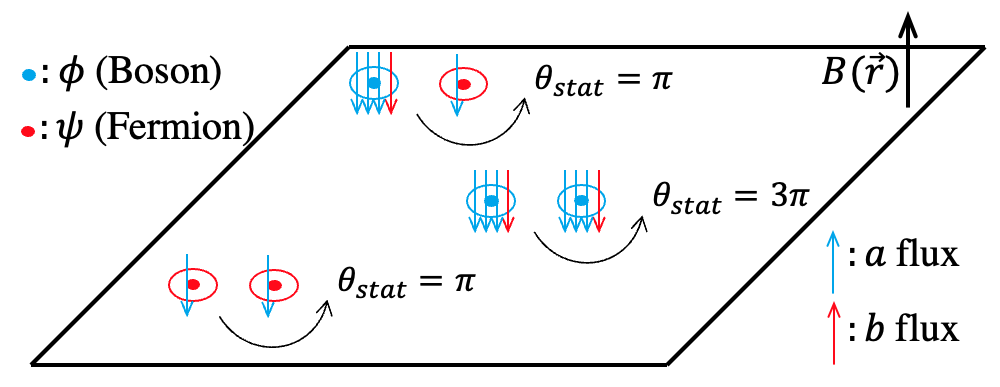}
    \caption{An illustration of the statistical angles between particle-flux composites introduced by the CS terms in Eq.~\ref{eq: CS terms}, which corresponds to $q=3$, $q'=0$ and $q''=1$ in the flux attachment matrix. }
    \label{fig: illustration}
\end{figure}

In this section, we present the two-fluid theory consisting of CBs described by the complex scalar fields $\phi$ and $\phi^\star$, CFs described by the Grassman fields $\psi$ and $\psi^\dagger$, and emergent dynamical gauge fields $a$ and $b$ in the presence of a background electromagnetic field $A$. Throughout this paper, we will adopt the following convention: Greek letters ($\mu,\nu,\dots$) are used for spacetime indices ($0,1,2$ represent $t,x,y$), and Latin letters ($i,j,\dots$) are used only for the spatial indices; the sign convention is such that $x^\mu = (t,x,y)$ and the metric $g_{\mu\nu} = \text{diag}(1,-1,-1)$. We absorb a factor of electron charge $|e|$ in the definition of $A$ and adopt units such that $\hbar = c = 1$. The Lagrangian density of the theory then reads:
\begin{align} \label{eq: Lagrangian}
    {\cal L} =&  \phi^\star \left[ i\partial_0 + a_0 + A_0 \right]\phi-\frac 1 {2m}\left| \left( -i\vec \nabla+ \vec a  + \vec A \right)\phi\right|^2  \nonumber \\ 
& +\psi^\dagger \left[ i\partial_0 + b_0 + A_0 \right]\psi -\frac 1 {2m}\left| \left( -i\vec \nabla+ \vec b + \vec A \right)\psi\right|^2\nonumber \\
&+\mathcal{L}_\text{CS} +  {\cal L}_\text{Int}\left[|\psi|^2,|\phi|^2\right] \\
\mathcal{L}_\text{CS}  =& - \frac{1}{4\pi}\begin{pmatrix}
a & b
\end{pmatrix} \mathcal{K}^{-1} \begin{pmatrix}
\mathrm{d}a \\ \mathrm{d}b
\end{pmatrix} \label{eq: CS terms}
\end{align}
where we have adopted a short-handed notation for exterior derivatives, \eg $a\mathrm{d}b = \epsilon^{\mu\nu\eta}a_\mu \partial_\nu b_\eta$, and
\begin{align}
\mathcal{K} \equiv \begin{pmatrix}
q & q'' \\
q'' & q'
\end{pmatrix}
\end{align}
enforces the flux-attachment constraints according to the equations of motion of $a$ and $b$:
\begin{align} \label{fluxattachment}
  \epsilon^{\mu\nu\sigma}  \begin{pmatrix}
   \partial_\nu a_\sigma \\ \partial_\nu b_\sigma
   \end{pmatrix} = 2\pi \mathcal{K}
\begin{pmatrix}
J^\mu_{\phi} \\
J^\mu_{\psi}
\end{pmatrix} 
\end{align}
where $J^0_\phi = \rho_\phi = |\phi|^2$, $J^0_\psi = \rho_\psi = |\psi|^2$ are the densities of CBs and CFs, and $J^i_\phi$, $J^i_\psi$ are the current vectors. $\mathcal{L}_{\text{Int}}\left[|\psi|^2,|\phi|^2\right]$ is a potential that represents the interactions among particles.  We divide the interactions into the sum of a short-range and a long-range piece, the former of which we effectively treat as a point-contact interaction, 
\begin{align}\label{eq: effective interaction}
    {\cal L}_\text{Int}[|\phi|^2, |\psi|^2] = &-  \left[ V_1 \rho_\phi^2 /2 + V_2 \rho_\phi \rho_\psi \right] \nonumber\\
    & \ \ \ \ \ \ -  V_\text{long range}[|\phi|^2+|\psi|^2]
\end{align}
where $V_1$ and $V_2$ are effective parameters. We note that due to Grassman algebra, $|\psi|^4$ vanishes such that the $\psi$ particles do not self-interact. 

With this flux attachment structure and $\phi$, $\psi$ being bosonic and fermionic fields, as illustrated in Fig.~\ref{fig: illustration}, one can see that the self statistics of $\phi$ and $\psi$ particles are fermionic for odd $q$ and even $q'$. We will restrict our choices of  $q$ and $q'$ accordingly, and regard the two components as two different flavors of electrons. The mutual statistical angle between the two flavors is determined by $q''$. In principle, this two-fluid theory should also include terms that can transmute $\phi$ and $\psi$ particles, \eg $\phi^\dagger \psi$, revealing their common underlying nature  - electrons. However, such terms cannot exist on their own and must be accompanied by monopole operators that globally rearrange the spacetime configuration of the CS gauge fields $a$ and $b$. (The nature of such terms is discussed in App.~\ref{CBCF}.) Such terms are thus unimportant as long as we are studying time-independent saddle-point solutions to the theory.

We expect this theory to effectively describe a {\it single-component} electron fluid in the presence of a background magnetic field $B \equiv -\epsilon^{ij}\partial_i A_j$. The electron density $\rho= \rho_\phi +\rho_\psi =|\phi|^2+|\psi|^2$ is thus a sum of the densities of the two artificial species.

\section{Uniform states}\label{sec: uniform}

We first analyze the filling fractions that are naturally described by this two-fluid theory. To derive those fractions, we note that the effective magnetic fields seen by $\phi$ and $\psi$ particles in a uniform system are:
\begin{align} \label{fieldcharge}
    B_\phi = & B - 2\pi(q\rho_\phi+q'' \rho_\psi) \\
    B_\psi = & B - 2\pi(q''\rho_\phi+q' \rho_\psi)
\end{align}
When $B_\phi = 0$, we expect the CBs to form a condensate, and when $B_\psi = 2\pi \rho_\psi / \nu_\text{CF}$ with $\nu_\text{CF}$ an integer, CFs to fill $|\nu_\text{CF}|$ effective LLs. Thus, the four integers $q,q',q'',\nu_\text{CF}$ specify a QH state. Specifically, our hybrid picture suggests the following wave-function ansatz for the electrons:
\begin{align}\label{eq: wavefunction}
    \Psi\left(z\right)& = \mathcal{A} \left[\prod_{i<j}\left(s_i-s_j\right)^{q}  \prod_{k<l} \left(w_k-w_l\right)^{q'}   \right.  \\
    & \ \ \ \ \ \  \left. \prod_{i,k} \left(s_i-w_k\right)^{q''} \Psi_\text{mix}\left(s,\bar{s};w,\bar{w}\right)\right] \mathrm{e}^{-\sum_i|z_i|^2/2}  \nonumber
\end{align}
where $s\equiv (s_1,\dots,s_{N_\phi})$ and $w\equiv (w_1,\dots,w_{N_\psi})$ are the complex coordinates of, respectively, the CBs and the CFs, $z\equiv (s_1,\dots,s_{N_\phi};w_1,\dots,w_{N_\psi})$ are the collection of {\it all} the coordinates, $\Psi_\text{mix}\left(s,\bar{s};w,\bar{w}\right)$ is a many-body wave-function of the mixture of CBs and CFs, $\mathcal{A}$ represents an anti-symetrization over all coordinates. In writing the wavefunction into this form we have assumed a symmetric gauge for the background field, and the complex coordinates are defined as $z_i = (x_i -\mathrm{i} y_i)/l_B$ where $l_B \equiv 1/\sqrt{B}$ is the magnetic length. Depending on the filling fraction, one may further enforce a projection of the wave function onto the lowest LL (LLL). For the current case, $\Psi_\text{mix}\left(s,\bar{s};w,\bar{w}\right) = \phi(s,\bar{s})\psi(w,\bar{w})$, where $\phi(s,\bar{s}) = 1 $ is a Bose condensate wavefunction~\cite{zhang1992chern,10.21468/SciPostPhys.8.5.079}, and $\psi(w,\bar{w})$ (modulo the final Gaussian factor) is a Slater determinant of CFs in $|\nu_\text{CF}|$ filled effective LLs. We will call a state defined in this way the $[q,q',q'',\nu_\text{CF}]$ state; its filling fraction is 
\begin{align}\label{eq: nu}
    \nu = \frac{2\pi\rho}{  B} = \frac{q+\tilde{q}'-2q''}{q\tilde{q}'-(q'')^2}
\end{align}
with $\tilde{q}'\equiv q'+1/\nu_\text{CF}$, while the density ratio between the two components is
\begin{align}
    \rho_\phi / \rho_\psi = (\tilde{q}'-q'') / (q-q''). 
\end{align}
It is necessary that $\nu>0$  and $ \rho_\phi/\rho_\psi\ge 0$; for $q''=$ even, there is an additional issue of whether the state vanishes upon anti-symmetrization, as discussed below.

Clearly, these states include many familiar ones. States with $\nu_\text{CF}=0$ are pure CB states that correspond to  Laughlin states at $\nu = 1/q$~\cite{PhysRevLett.50.1395}. States with $q''=\tilde q'$ are pure CF states, which is possible only if $|\nu_\text{CF}|=1$ or $\infty$, given the integer constraint of $q'$, $q''$ and $\nu_{\rm CF}$~\footnote{Since $\nu_\text{CF}$ cannot be other integers, not all the states in the Jain sequence~\cite{jain2007composite} can be accessed with pure CFs within this framework. }. For $\nu_\text{CF}=\infty$ they correspond to a composite Fermi liquid (CFL) with $\nu = 1/q'$, while for $\nu_{\rm CF}=\pm 1$ they are a CF version of the Laughlin state at $\nu = 1/(q'\pm 1)$.

Turning to multi-component states (which have been discussed within pure CB or CF approaches~\cite{PhysRevLett.84.3950, PhysRevB.66.155302, PhysRevB.51.4347,PhysRevLett.127.246803}), note that states with $\nu_\text{CF}=\pm 1$ and  $q''\neq \tilde q'=q$ have equal densities of CFs and CBs and the same filling fraction and Hall responses as $(q,q,q'')$ Halperin states~\cite{Halperin:1983zz},  the {\it anti-symmetrized version} of which are in turn related to various hierarchical states~\cite{10.21468/SciPostPhys.8.5.079} and Read-Rezayi states~\cite{PhysRevB.81.045323}.  The simplest example is $[3,2, 2,1]$ which has $\nu=2/5$ like the $(3,3,2)$ Halperin state. The identification between these two states is precise if the CFs are taken to occupy the lowest Landau level.  Since when $q''$ is even, the $(q,q,q'')$ Halperin states vanish upon anti-symmetrization, so do the corresponding $[q,q\mp 1,q'',
\pm 1]$ states; indeed, as pointed out in Ref.~\cite{10.21468/SciPostPhys.8.5.079}, in order to give a correct wavefunction for this sort of hierarchical state, an additional degree of freedom - ``orbital spin'' - should be included in order to differentiate the wavefunctions of the two components.  Within our framework, this could be conveniently achieved by considering a state in which the CFs fill the first effective LL and leave the zeroth effective LL empty, as in the conventional CF construction.

One of the most interesting new possibilities revealed by this approach is  a coexistence of a CB condensate and a CFL ($[q,q', q''\neq q', \nu_\text{CF}=\infty]$).  Note that, in this case, the Fermi sea volume of the compressible CF state times the spatial area per flux quantum is not equal to the filling fraction $\nu$, signaling a  deviation from the Luttinger relation~\cite{PhysRev.119.1153}. Similar physics have been proposed in the context of doped spin liquids - when part of the electrons form a topologically ordered state while the remaining part forms a Fermi liquid, such states of matter have been dubbed ``Fermi Liquid$^*$'' (FL$^*$)~\cite{PhysRevLett.90.216403}. In the same spirit, we name the coexisting phases of CB condensate and a CFL ``composite Fermi liquid$^*$'' (CFL$^*$). To give an example, $[1,0, 2, \infty]$ is such a state describing $\nu=3/4$, in which $2/3$ of the electrons form a condensate while $1/3$ of the electrons form a CFL. 

We mention that similar trial uniform wavefunctions (but not field theory formalisms) that have coexisting CBs and CFs have been proposed in Refs.~\cite{PhysRevLett.91.046803,PhysRevB.79.125106}, where they were used to describe the transition between a CFL state and a CB condensate state in a QH bi-layer.

\section{Quasi-particle in Laughlin states} \label{sec: quasiparticles}

In Ref.~\cite{PhysRevLett.62.82}, Zhang, Hansson, and Kivelson described the Laughlin FQH liquids at $\nu=1/q$ using the CB part of Eq.~\eqref{eq: Lagrangian}, \ie the CSGL theory. In addition to the correct FQH response, the saddle-point treatment of this theory provides natural vortex solutions that describe the fractionally charged anyonic quasiholes. However, the theory also has deficiencies---among other things, it does not simply capture the intra-LL magneto-roton spectrum at small momentum. Also, the natural candidate for a quasielectron, namely a fundamental anti-vortex, is unappealing in that the charge density vanishes at the center of the quasiparticle~\cite{TAFELMAYER1993386}.  In this section, we show how the extension of the CSGL theory to our two-fluid theory allows for more general and physically appealing quasiparticle solutions (including quasielectrons and neutral excitations), which are the bound states of vortices of the CB condensate and several CFs. Exploiting the freedom to choose different values of $q'$ and $q''$ (as long as $q'$ is even), we find distinct soliton solutions of the saddle-point equations having the lowest energy for different ranges of interaction strength, all of which have the same charge and orbital spin, but different quadrupole moments. We present the main results here and defer the detailed discussions to App.~\ref{app: sec: excitations}.

\subsection{Formal considerations}

We first consider the topological properties of the possible soliton excitations. In App.~\ref{app: sec: formal} we derive that, for a soliton bound state made from  $n_v$ vortices and $n_\psi$ CFs (with both $n_v$ and $n_\psi$ integers), the charge $Q$, self-statistical angle $\theta_\text{stat}$, and orbital spin $S^z$ (for rotationally invariant systems) are given by,
\begin{align}
    Q &= -n_\psi + \frac{n_v+q''n_\psi}{q} \label{chargeconstraint} \\
    \theta_\text{stat} &= q \pi Q^2  \mod 2\pi \\
    S^z
    & = L_\psi^z +  \frac{(q'-q) n_\psi - qQ }{2},
\end{align}
where $L_\psi^z $ is the angular momentum carried by the CFs. The self-statistical angle is independent of the sign of the charge and satisfies a familiar charge-statistics relation.

\begin{figure*}[t!]
\subfigure[]{\label{q3Q-1E}\includegraphics[width=0.32\linewidth]{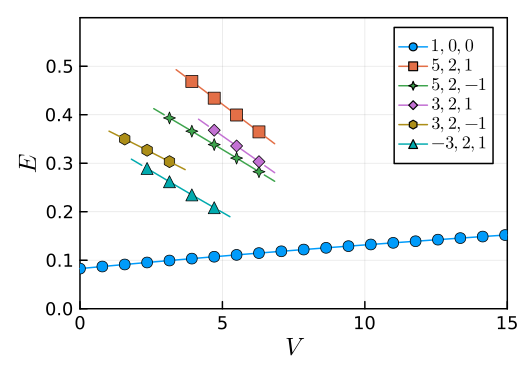}}
\subfigure[]{\label{q3Q0E}\includegraphics[width=0.32\linewidth]{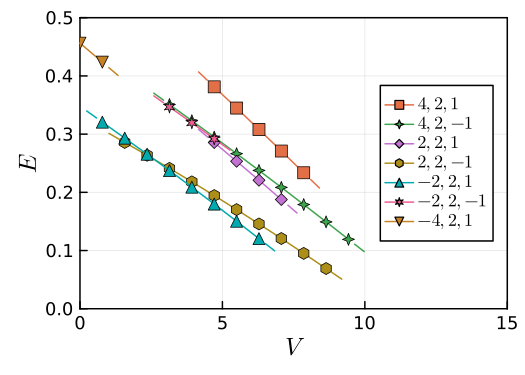}}
\subfigure[]{\label{q3Q1E}\includegraphics[width=0.32\linewidth]{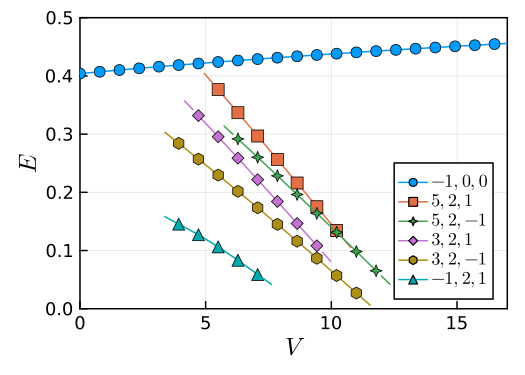}}\\
\subfigure[]{\label{q3Q-1Q}\includegraphics[width=0.32\linewidth]{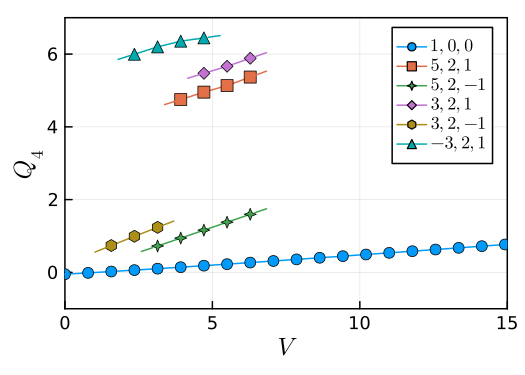}}
\subfigure[]{\label{q3Q0Q}\includegraphics[width=0.32\linewidth]{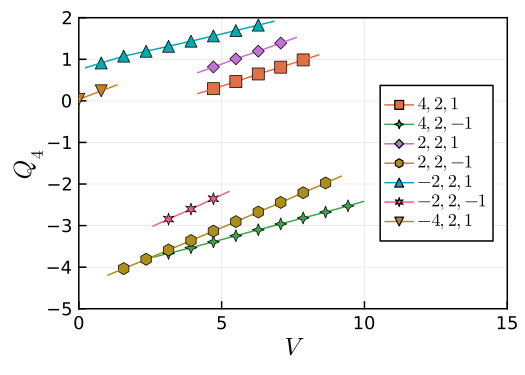}}
\subfigure[]{\label{q3Q1Q}\includegraphics[width=0.32\linewidth]{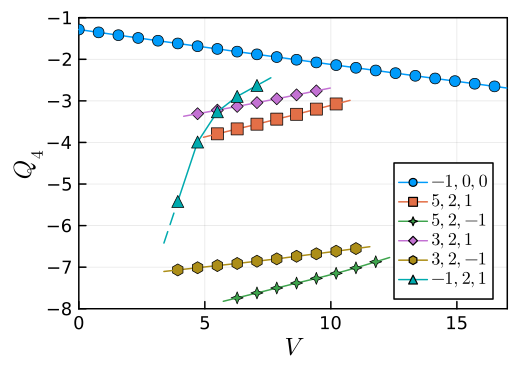}}
\caption{\label{qQplots} For a $\nu=1/q=1/3$ FQH state described by Laughlin state, as functions of the (short-range) interaction strength $V=V_2=V_1$, we compute the energy $E$ and the quadruple moment $Q_4$ of different soliton solutions with charge $Q=1/3$ (quasi-hole) in panels a and d,  $Q=0$ (neutral excitations) in panels b and e,  and $Q=-1/3$ (quasielectron) in panels c and f.  $E$, $V$, $Q_4$ are repectively measured in units of cyclotron frequency $\omega_c = B/m$, inverse mass $1/m$, and squared magnetic length $l_B^2 = 1/B$. The label represents the value of $n_v,n_\psi,L^z_\psi$ (vorticity, CF number, and CF angular momentum), and the corresponding values of $q',q''$ can be inferred from Eqs.~\ref{chargeconstraint} and \ref{Lzconstraint}. The curves are plotted in the range of interaction strength where the solitons are stable. Except for the vortex and the anti-vortex ($1,0,0$ and $-1,0,0$) which have been studied in Ref.~\cite{TAFELMAYER1993386}, all other solutions plotted have $n_\psi=2$, \ie two CFs are bounded to the vortices.}
\end{figure*}

With these general topological properties of the excitations, we may constrain the composition of the solitons that are of interest.  In this work, for concreteness, we will investigate the quasi-holes, quasielectrons, as well as certain neutral excitations. On physical grounds, we have only explicitly considered quasiparticles that satisfy the topological non-relativistic spin-statistics relation~\cite{balachandran1993spin}
\begin{align} \label{spincharge}
     q Q =  - 2 S^z
\end{align}
which leads to the constraint
\begin{align}
(q-q')n_\psi &= 2 L_\psi^z \, .\label{Lzconstraint}
\end{align}

\subsection{Results}

We have carried out explicit calculations with $q=3$ (\ie for the $\nu=1/3$ Laughlin state), and looked for soliton solutions for various compositions and different values of $q'$ and $q''$ consistent with Eq.~\ref{Lzconstraint}. To be concrete, we have fixed $V_1 = V_2 = V$ and (since the solitons are relatively small) have neglected the long-range part in the effective interaction in Eq.~\ref{eq: effective interaction}. Then, we numerically solve the saddle-point equations $\frac{\delta \mathcal{L}}{\delta \phi}, \frac{\delta \mathcal{L}}{\delta \psi} = 0$ (see Appendix.~\ref{app: sec: saddlepoint} for details). For simplicity, we assume rotational symmetry of the soliton, such that only the radial dependence of the field amplitudes needs to be accounted for. In order to self-consistently solve these coupled saddle-point equations, we performed a numerical calculation with a relaxation algorithm on a radial coordinate discretized into at least $5000$ mesh points for $r/l_B\in [0,12]$. Each solution we obtained was verified to be convergent to a relative precision of $10^{-2}$ in the value of the energy.  Due to limited computational resources, we only considered cases with ``small'' soliton solutions that is with $|n_v|\le 5$ and $n_\psi \le 2$.

Indeed, we found multiple solutions for each case of fractional charge $Q=\pm 1/3$ and $0$. As shown in Fig.~\ref{qQplots}, those solutions have distinct energies $E$ and quadruple moments $Q_4 \equiv - \int d^2 \Vec{r} r^2 [\rho(r) - \Bar{\rho}]$, and are stable in different ranges of the interaction strengths $V$. These observations suggest that, depending on the microscopic details of the interactions, the nature of the lowest energy excitations may be entirely modified without any changes in the ground state properties.

\begin{figure}
    \centering
    {\includegraphics[width=\linewidth]{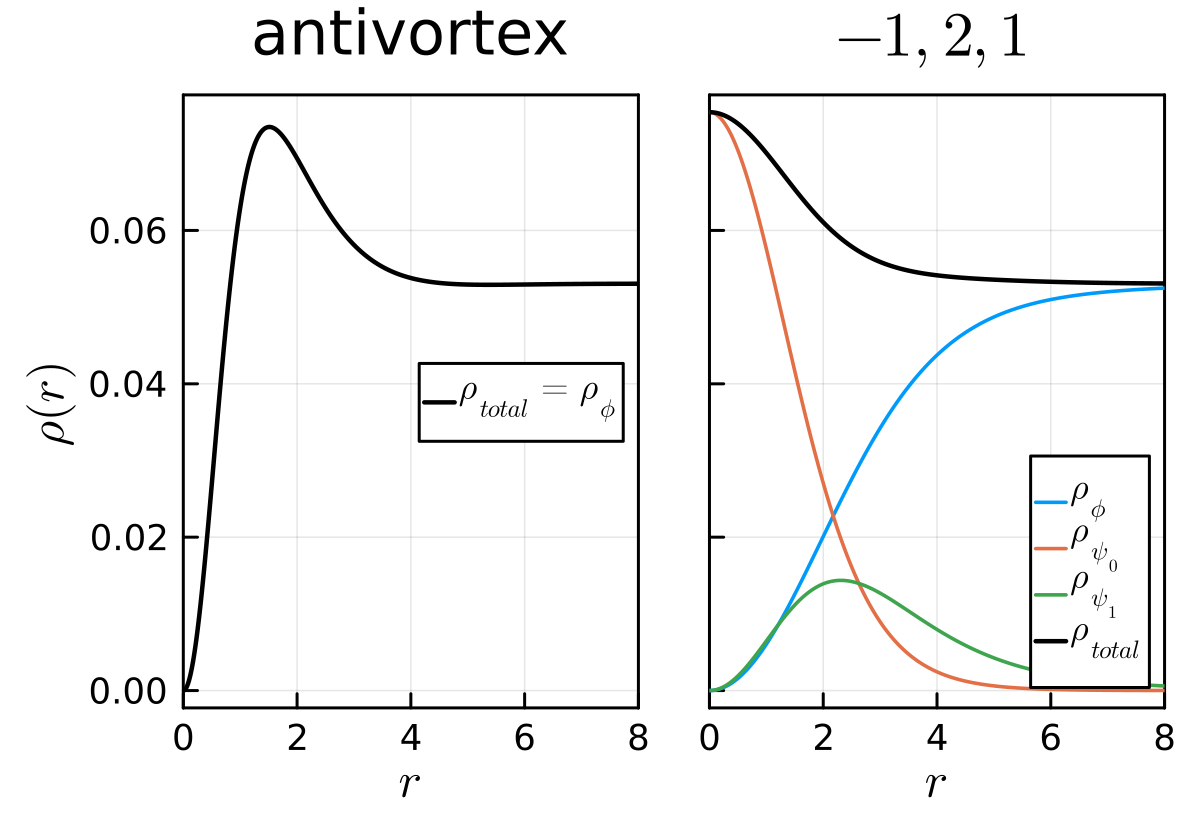}} 
    \caption{\label{ACprofile} Representative density profiles $\rho(r)$ of quasielectron excitation with $Q=-1/3$ out of a $\nu=1/3$ FQH state described by a Laughlin state, with (left panel) the anti-vortex solution, and (right panel) the composite solution with $n_v, n_\psi, L^z_\psi = -1, 2, 1$ and $q,q',q''=3,2,3$. Both profiles are obtained at $mV\approx 5.5$, and magnetic length $l_B = \sqrt{1/B}$ is set to $1$. In (b), $\psi_0$ and $\psi_1$ represents the CF states with angular momentum $L^z = 0,1$ respectively. }
\end{figure}

As one may expect, the pure CB vortex solution has the lowest energy among the solutions for quasi-holes. By contrast, for quasielectrons, there are various composite solutions consisting of CB vortices and CFs which, for an intermediate range of interaction strengths, have lower energies than the anti-vortex solution. These composite solutions also have more realistic density profiles than the naive anti-vortex solution: As shown in Fig.~\ref{ACprofile}, the density $\rho(r)$ of the bare anti-vortex solution vanishes at the origin, has a large density modulation extending over a wide range of $r$. In contrast, $\rho(r)$ for a composite solution is much smoother.  

Especially, we note that, in a wide range around $m V\approx 5$, the most stable quasielectron solution ($n_v, n_\psi, L^z_\psi = -1, 2, 1$) is given by a flux attachment matrix with $q'=2$, and can be viewed as a bound state of $-1$ vortices of $\nu=1/3$ FQH fluids and $2$ CFs at $\nu=1/2$. This fact hints at the ``naturalness'' of these composite solutions. Eventually, the energetic advantage of these types of novel composite solutions in our simplified calculation (at least for those with positive $n_v$) suggests that in future serious numerical simulations with realistic setups, one should consider the more general class of wave functions Eq.~\ref{eq: wavefunction} with $\Psi_\text{mix}(s,\bar{s};w,\bar{w}) = \phi(s,\bar{s}) \psi(w,\bar{w})$ where $\psi(w,\bar{w})$ is a few-body fermion wavefunction to be optimized in subsequent variational calculations, and $ \phi(s,\bar{s}) = \prod_i (s_i-\eta)^{n_v} $.

Remarkably, besides novel quasielectron solutions, we also found localized neutral excitations that have not yet been theoretically predicted. It may be an exciting possibility that these might describe the intra-LL magneto-roton excitations. We find that, in the regime where such excitation is stabilized, its energy is less than the sum of the energies of a quasielectron and a quasi-hole, indicating that this may be a Girvin-MacDonald-Plazman mode distinguishable from the (quasi-)particle-hole continuum at zero momentum.

We now compare our proposal to existing descriptions of the quasielectron. In the CF picture, a quasielectron is an extra CF in the next effective LL. This is a simple picture, but it is not immediately evident that this excitation is a fractionally charged anyon. In the CB picture, this is clear since an effective anti-vortex has the correct charge and topological properties. However, as already mentioned, the charge distribution of an elementary anti-vortex is unphysical, an issue that does not arise for the energetically favored vortex.

The quantity $S^z$ is a localized orbital angular momentum, or “orbital spin”, which is unrelated to the fundamental spin of the electrons. In numerical calculations, the orbital spin can be deduced from the relation between the number of particles and the number of flux quanta for the ground state on a sphere and is consistent with the value predicted by the spin-charge relation \eqref{spincharge}. For the Laughlin quasiparticles discussed here, $S^z = \pm 1/2$ so in the CF picture the difference between the quasihole and the quasielectron is naturally explained since the latter has a single CF in the first effective LL, which involves an extra unit of $L^z$ compared to the states in LLL. In the CB picture, the orbital spin can be interpreted as the angular momentum of a charge-vortex bound state, which is most easily explained in terms of a descendant vertex operator in a conformal field theory~\cite{PhysRevB.76.075347}. It is quite pleasing that the preferred quasielectron solution discussed above naturally fits in the CFT description. Details on this are in App.~\ref{app: sec: CFT}. We also note that there are other approaches to quasielectron wave functions, as \eg in Ref.~\cite{bochniak2022mechanism}. To summarize: Although the composite solutions for quasielectron in our formalism seem complicated, we stress that they require fewer {\it ad hoc} assumptions compared to existing schemes. They do not rely on any concept of ``effective Landau level'' as in CF theories, nor an artificially introduced orbital spin as in an effective CB theory~\cite{10.21468/SciPostPhys.8.5.079}.

\section{QH Interfaces}\label{sec: interface}

In this section, we apply our theory to a smooth interface $B(x)$  between two regions with different magnetic fields. Far to the left, $B(\vec r)\to B_- =2\pi q \Bar{\rho}$, at which a $\nu =1/q$ FQH state (with $\nu_\text{CF} = 0$) is stable, while far to the right, $B(\vec r)\to B_+$,  at which there is the compressible state $[q,q',q'',\nu_{\text{CF}}=\infty]$ with $\nu$ given in Eq.~\ref{eq: nu}.  For instance, this includes the case in which $q''=q'$, $B_+=(q'/q) B_-$, and the state at $x\to\infty$ is a CFL with $\nu=1/q'$; for $q''\neq q'$ the compressible state is a CFL$^*$. In all cases, we assume that we can take a single value of $q$, $q'$, and $q''$ in describing the state of the system for {\it all positions}. We further assume the electron density is constant $\rho(\vec r)=\Bar{\rho}$ when viewing the system at distances larger than a scale $\ell_\rho$, which is a presumed consequence of the presence of long-range Coulomb interactions and the condition of charge neutrality. In our treatment, we will impose $\rho = \bar\rho$ as a constraint.

We will consider smooth $B(\vec{r})$ profiles, which vary over a length scale $L$ that is greater than all the microscopic length scales
\begin{align}\label{eq: large interface}
    L \gg l_B, \ell_\rho, \ell_\text{mf}
\end{align}
where  $l_B$ is the lager magnetic length for $B=B_\pm$ and $\ell_\text{mf}$ is a CF mean free path discussed below.

Due to disorder broadening, the CF compressibility, $\kappa$, can be taken to be approximately constant. Therefore, we use a local expression for the CF kinetic energy,
\begin{align}\label{eq: CF kinetic energy}
    \mathcal{E}_\text{kin} = \rho_\psi^2/(2\kappa)
\end{align}
This term thus acts as a potential, and effectively provides a ``Fermi degenerate pressure''. Meanwhile, due to disorder scattering, we assume that the CFs are dissipative and thus cannot support any persisting current,
\begin{align}
    J_\psi^i =0, \label{eq: zero current}
\end{align}
when considering phenomena at length scales above the mean free path $\ell_\text{mf}$ of the CFs. We note that both assumptions about the physics of the CFs can be improved, \eg by introducing more sophisticated local response terms. 

The goal of this section is to derive the interpolating behavior between the two regions, and especially determine how the CB component $\rho_\phi$ changes over the interface. A detailed derivation for a representative case can be found in App.~\ref{app: sec: interface}. 

We start our analysis by adopting the dual representation of CB part of the theory. To do so, we parametrize $J_\phi^\mu =\epsilon^{\mu\nu\sigma} \partial_\nu h_\sigma/(2\pi)$ with a hydrodynamic field $h$ and rewrite the CB Lagrangian in \eqref{eq: Lagrangian} as
\begin{align}
   \mathcal{L}_{\phi}[h,a] & = \frac{1}{2\pi}(a+A)\d h + \frac{m \vec{J}_\phi^2}{2 \rho_\phi} - \frac{(\nabla \rho_\phi)^2}{8 m\rho_\phi} \ .
\end{align}
We note that by rewriting the theory into this form, we have assumed that the CBs remain condensed and vortices are expelled so that the quasi-particle current which couples to $h$ can be neglected. 

Next, we treat the CF part of the theory in an effective manner. The constraints on the density will be imposed by a Lagrangian multiplier field $\eta$, which can be interpreted as representing the effects of the long-ranged interactions. Including the effective kinetic energy in Eq.~\ref{eq: CF kinetic energy} and incorporating the zero current condition in Eq.~\ref{eq: zero current}, we have written the effective Lagrangian for the CFs as,
\begin{align}
    \mathcal{L}_{\psi,\text{eff}}[\rho_\psi; b;\eta]  =& \rho_\psi (b_0 + A_0) - \frac{\rho_\psi^2}{2\kappa } \nonumber\\
    & \ \ \ + \eta (\rho_\phi+\rho_\psi-\bar{\rho}) \ .
\end{align}

We will seek steady-state solutions for spatially non-uniform magnetic fields $B(\vec{r})$ and with no electric field. We then pick a static gauge, $h (\vec x,t) = h (\vec x)$ for the hydrodynamic field, statistical gauge field $b$, and background field, such that all terms containing time derivatives can be dropped from the Lagrangian.  Finally, we shift $b\rightarrow b+q'' h$ to obtain an effective Lagrangian:
\begin{align}
    &\mathcal{L}_\text{eff}[\rho_\psi; h,b;\eta]\nonumber\\ 
    =&   \rho_\psi (q'' h_0 + b_0 + A_0) - \frac{\rho_\psi^2}{2\kappa } + \eta (\rho_\phi+\rho_\psi-\bar{\rho}) \nonumber\\
    & -\frac{1}{4\pi q'} b \d b  -\frac{V_1}{2} \rho_\phi^2  -V_2 \rho_\phi \rho_\psi + \frac{m(\nabla h_0)^2}{2 (2\pi)^2 \rho_\phi} \nonumber\\
    & \     - \frac{(\nabla \rho_\phi)^2}{8 m \rho_\phi} +\frac{q}{4\pi} h\d h + \frac{1}{2\pi} A \d h  \label{eq: effectiveLagrangian}
\end{align}
where $\rho_\phi = -\frac{B_h}{2\pi} \equiv \frac{1}{2\pi} (\partial_1 h_2-\partial_2 h_1)$ should be understood. We see that $b$ still attaches $q'$ flux to each CF. 

In App.~\ref{app: sec: interface}, we investigate the full set of saddle-point equations, but the ones obtained by varying $h$ and $\rho_\psi$ are particularly useful,
\begin{align} 
  \frac{\delta \mathcal{L}}{\delta h_0}  &: \ \ \  q''\bar{\rho} + (q-q'')\rho_\phi \nonumber\\
     & \ \ \ \ \ \ \ \ - \frac{B}{2\pi} - \frac{m}{(2\pi)^2} \nabla(\frac{1}{\rho_\phi}\nabla h_0) = 0\label{eq: h0EOM}\\
   \frac{\delta \mathcal{L}}{\delta h_i} &: \ \ \ \epsilon^{ij} \partial_j  \left\{ qh_0 +\eta - V_1\rho_\phi - V_2 \rho_\psi - \frac{m^2(\nabla h_0)^2}{2(2\pi)^2 \rho_\phi^2} \right.\nonumber\\
    & \ \ \ \ \ \ \ \ \ \ 
    \left. +\frac{1}{8 m } \left[ 2\nabla( \frac{1}{\rho_\phi}\nabla \rho_\phi) + \frac{(\nabla \rho_\phi)^2}{\rho_\phi^2} \right]  \right\}  = 0 \label{eq: hiEOM} \\
     \frac{\delta \mathcal{L}}{\delta \rho_\psi} &:\ \ \ q'' h_0 +\eta -\rho_\psi /\kappa - V_2 \rho_\phi =0 \label{eq: rhopsiEOM}
\end{align}
where we have used the constant density constraint as well as the other equations of motion. The physical meaning of Eq.~\ref{eq: h0EOM} is that the local net magnetic field seen by the CBs, $B_\phi=B- 2\pi(q\rho_\phi + q''\rho_\psi)$, induces a CB current; 
Eq.~\ref{eq: hiEOM} means that the net potential experienced by CBs modulates the density profile; Eq.~\ref{eq: rhopsiEOM} implies that the net force on the CFs is zero. From Eqs.~\ref{eq: rhopsiEOM} and \ref{eq: hiEOM}, we see that $\eta$ indeed can be interpreted as an electric potential seen by both $\phi$ and $\psi$ provided by the long-range interactions.

For simplicity, from here on,
we assume that the magnetic field strength only depends on $x$, \ie the interface is along $y$ axis. 

As already mentioned, we will show that the CB condensate extends into the interface. To study this ``proximity" effect, we define 
\begin{align}
    &\delta \rho(x) \equiv  \rho_\phi(x) - \rho_B(x) \\
    &\rho_B(x)\equiv \frac{\left[
    {B(x)}/{2\pi} -q''\bar{\rho}\right]}{[q-q'']}\ . \label{eq: rhoB definition}
\end{align}
This parametrization is such that if  $\delta\rho(x)=0$, the effective magnetic field seen by the $\phi$ particles, $B_\phi$, would vanish. 

Integrating Eq.~\ref{eq: h0EOM}, we obtain a {\it formal} solution for $h_0(x)$ as a functional of $\rho_\phi(x)$:
\begin{align}\label{eq: h0 formal solution}
    h_0(x) =& (q-q'') \frac {(2\pi)^2} m \int_{-\infty}^{x}  \mathrm{d}x'  \rho_\phi(x') \int_{-\infty}^{x'} \mathrm{d} x'' \delta\rho(x'') 
\end{align}

Combining Eqs.~\ref{eq: hiEOM} and \ref{eq: rhopsiEOM}, and substituting in the above formal solution Eq.~\ref{eq: h0 formal solution}, gives an integro-differential equation,
\begin{align}\label{eq: integro differential}
    &(q-q'')^2(2\pi)^2 \int_{-\infty}^{x} \mathrm{d}x' \rho_B(x') \int_{-\infty}^{x''} \mathrm{d} x'' \delta\rho(x'')  \nonumber\\
    = & m\tilde{V}\left[\bar{\rho}-\rho_\phi(x)\right] - \frac{1}{4} 
    \left[\frac{\ddot \rho_\phi}{\rho_\phi} - \frac{(\dot \rho_\phi)^2}{2\rho_\phi^2}\right]_{x} \, .
\end{align}
where we define $\tilde{V} \equiv 2V_2-V_1-1/\kappa$.

This equation is in general hard to solve analytically. However, an approximate solution (which we will compare with the exact numerical solution later) can be obtained as follows:
\begin{itemize}
    \item Noting that all the fields should vary at scale $L\gg \ell_B \sim 1/\sqrt{\rho}$, we drop the terms in Eq. \ref{eq: integro differential} that involve explicit spatial derivatives.
    \item Since we already assumed CB remains condensed with no quasiparticles, the net field seen by the CBs should be small compared to 
    the CB density, \ie  $|\delta \rho| \ll \rho_\phi$, so we can approximate explicit factors of $\rho_\phi$ by $\rho_B$.
\end{itemize}
These approximations, which we will more carefully justify later, allow us to solve for $\delta \rho(x)$ and $h_0 (x)$ analytically:
\begin{align} \label{eq: delta rho solution}
    \delta \rho(x) &\approx - \frac{m\tilde{V}}{(q-q
    '')^2(2\pi)^2} \frac{\mathrm{d}^2}{\mathrm{d}x^2 } \left[\ln \rho_B (x) \right]\\
    h_0 (x) &\approx \frac{\tilde{V}}{(q-q'')} \left[\bar{\rho}- \rho_{B}(x)\right]\ . \label{eq: h0 solution}
\end{align}
Since $J^i_\phi = \epsilon^{ij} \frac{\partial_j h_0}{2\pi }$, the total current along the interface can be computed as well:
\begin{align}\label{eq: current solution}
 I_y = \frac{h_0(-\infty) - h_0(\infty)}{2\pi} = \frac{\tilde{V} (B_+-B_-)}{[2\pi (q-q'')]^2}
\end{align}
Eqs.~\ref{eq: delta rho solution}-\ref{eq: current solution} are the central results of this section, which suggests that for the setups and assumptions we are taking, the density profile of the CB condensate, $\rho_\phi(x)$, follows $\rho_B(x)$; the deviation $\delta\rho(x)$ is controlled by a phenomenological parameter $m\Tilde{V}$, and is small as long as the magnetic field is slowly varying. This result suggests that the CB condensate may well ``penetrate'' into the interface region where it coexists with the metallic component, which is the QH proximity effect we alluded to earlier. Heuristically, this effect further leads to a nearly quantized QH response in an essentially metallic regime. 

We now examine more carefully the justification for the two approximations used to solve  Eq.~\ref{eq: integro differential}. To safely neglect the spatial derivatives, it is necessary that $|m\tilde{V}|(\Bar{\rho}-\rho_\phi) \gg 1/L^2$. In order to safely approximate $\rho_\phi$ with $\rho_B$, we need $|\delta\rho| \ll \rho_\phi$, which, according to Eq.~\ref{eq: delta rho solution}, implies $|m\tilde{V}|/L^2 \ll \rho_\phi$. Thus the approximations in obtaining the solution are justified as long as
\begin{align} \label{eq: mV conditions}
    \frac{1}{L^2 (\bar{\rho} - \rho_\phi)} \ll |m\tilde{V}| \ll L^2 \rho_\phi
\end{align}
Given $L\gg \ell_B$, this is always satisfied as long as $\rho_\phi$ and $ \rho_\psi = \bar{\rho} - \rho_\phi$ are not too small compared to $\bar{\rho}$. It breaks down far enough into the pure CB or CF regime, but for large $L$ it holds over a broad intermediate range of $x$ on the interface.

\begin{figure}
    \centering
    \includegraphics[width = 0.9 \linewidth]{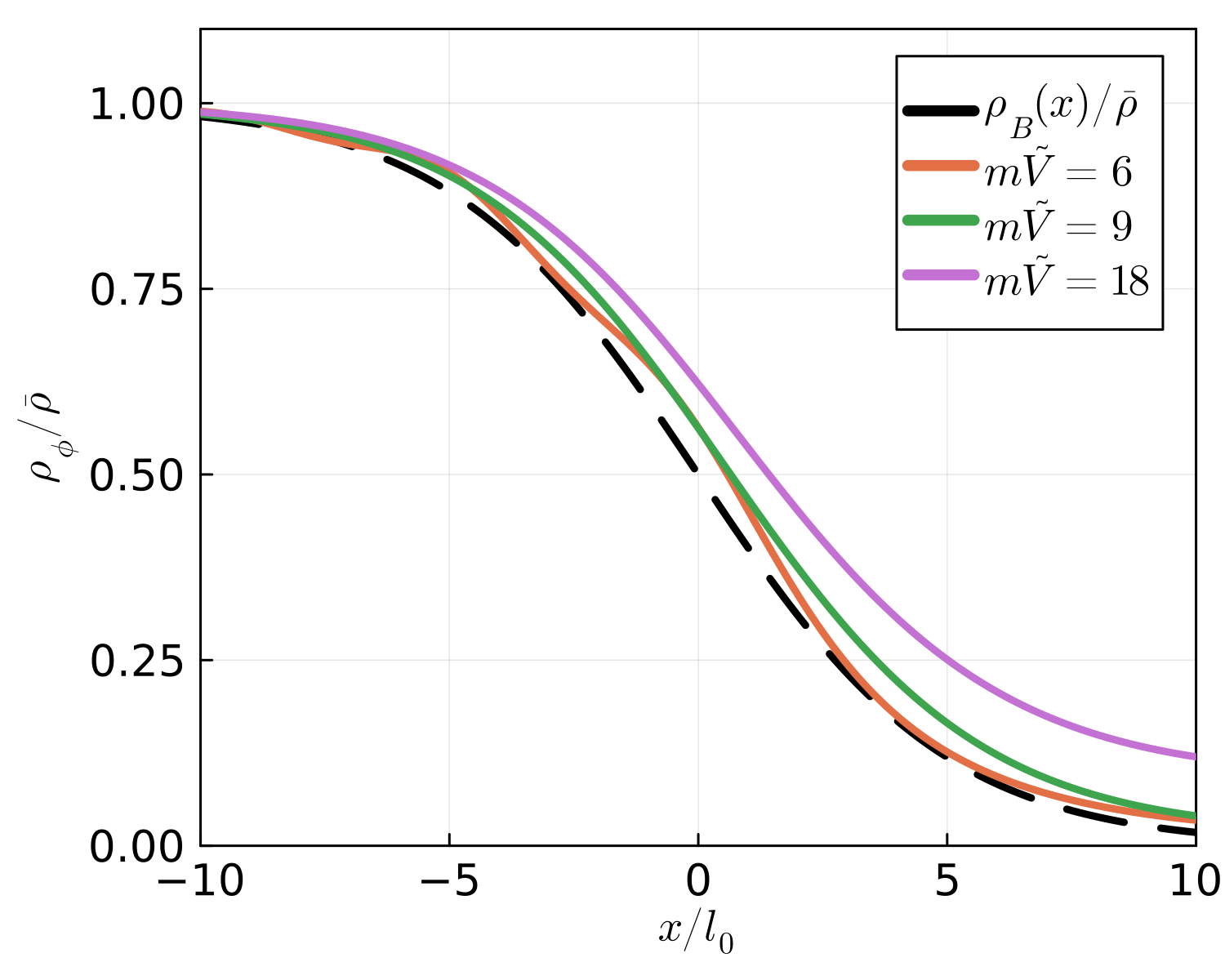}
    \caption{ The density profile $\rho_\phi(x)/\bar{\rho}$ numerically solved from Eq.~\ref{eq: integro differential} with $q=1$ and $q'=q''=0$ for magnetic field profile $B(x)= B_0 \left(1-\tanh\frac{x}{L}\right)/2$ with $L=5l_0$, where $l_0=1/\sqrt{B_0}$ and $\bar \rho=B_0/(2\pi )$ are respectively the magnetic length deep in the QH regime $x\rightarrow -\infty$ and the electron density. The black dashed line is $B(x)/B_0$. 
    }
    \label{fig: densityprofile_q1}
\end{figure}

To test the qualitative correctness of the solution in Eq.~\ref{eq: delta rho solution}, we numerically solved Eq.~\ref{eq: integro differential} for various $m\Tilde{V}$ with $q=1$, $q''=q'=0$, which corresponds to a $\nu=1$ QH to a $B=0$ metal interface. [In Fig.~\ref{fig: densityprofile_q3} of App.~\ref{app: sec: interface} we show the analagous results for $q=3$.] Indeed, for a magnetic field configuration with a {\it finite} interface width (which in  Fig.~\ref{fig: densityprofile_q1} we have taken to be $L= 5 l_0$ where $l_0$ is the magnetic length deep in the QH regime), we find that $\rho_\phi(x)$ roughly follows $\rho_B(x)$ (\ie the properly scaled magnitude of $B(x)$ in this case) with a deviation that is roughly proportional to $m\Tilde{V}$ (which can be seen by comparing the results with $m\Tilde{V} = 9$ and $18$).  Moreover, as expected, where the inequality in Eq.~\ref{eq: mV conditions} is not satisfied, extra features exist in the numerical solution: i) Especially for larger values of $m\tilde V$, $\rho_\phi$ decreases toward $0$ much more slowly than does $B(x)$ deep in the metallic regime. However, it should be noted that there is no reason for our theory to apply in the regime where $\rho_\phi$ is small, so the tail behavior is not physically meaningful. ii) There is a small oscillatory component visible in the case of a small $m\tilde{V}=6$ result with an onset at the edge of the pure CB regime. This piece becomes progressively more pronounced for still smaller values of $m\tilde{V}$, and is a potentially physically interesting effect that is not captured by the approximate solution from Eq. \ref{eq: delta rho solution}.

Slightly generalizing the theory in this section, the current formalism may also be used to study QH-QH interfaces at different filling fractions, which are specified by the same $q,q',q''$  but two different integer fillings for CF. One may further consider an FQH-Metal-FQH junction with a relatively narrow metal region with $B$ not reaching the value corresponding to the metallic state. Then, based on the above analyses, the CB condensate density $\rho_\phi$ remains finite across the junction, so that its phase coherence and QH response might be maintained despite the large variation in the magnetic field strength across the junction.

\section{Discussions}\label{sec: discussion}

The physical excitations in the compressible phase of a half-filled LL (and related phases) are not quite quasi-particles~\footnote{A composite Fermi liquid provides a sufficiently accurate cartoon of the state at $\nu=1/2$ to account for most of the observed behaviors. 
However, in the context of the CF theory, gauge fluctuations are sufficiently violent as to preclude the existence of well-defined quasi-particles in any conventional sense~\cite{PhysRevB.47.7312}.}, but there is nonetheless no doubt that the correct starting description is in terms of Fermi-surface excitations of a CFL.  
An FQH liquid can be described either in terms of filled LLs of CFs or in terms of the CSGL theory of CBs.  The latter provides a more direct description of the essential physics in terms of a precise correspondence with superconductivity --- the quantized Hall conductance comes from the ability of the condensate to carry a dissipationless current; the incompressibility from the Meissner effect; and the charge and statistics of the quasi-particles from flux quantization~\cite{kivelson1996electrons}. 
Conversely, it has been shown that an approach based on CF wave functions can account, with remarkable quantitative precision, for much of the low-energy excitation spectrum --- something that is generally beyond the scope of any field-theoretic approach.  In particular, a simple treatment of the short-scale structure of the quasielectron and of the collective mode spectrum has been lacking from CSGL perspective.

In general, a field theory can only be expected to capture the long-distance properties of a physical system, and as such, describe the system close to a critical point if not elsewhere. A compelling (and at least partially successful) account of QH plateau transitions in the presence of disorder has been achieved from the CB perspective by analogy with the magnetic field-driven superconductor-to-insulator transition~\cite{PhysRevB.46.2223}, or from the field theories with a CF perspective~\cite{doi:10.1146/annurev-conmatphys-033117-054227, PhysRevResearch.4.033146}. Theoretically, consistent formulations of the QH to QH nematic phase transition based on CSGL~\cite{PhysRevB.88.125137,PhysRevB.84.195124} or other field theories~\cite{PhysRevX.4.041050, golkar2016spectral, PhysRevX.7.041032} have been constructed, where the transition is associated with the softening and condensation of a $\vec q=\vec 0$ quadrupolar mode.  
A similar approach can also describe the transition from a fractional Chern insulator to a regular band insulator.

In general, the existence of a ``web of dualities''~\cite{SEIBERG2016395} allows any given problem to be described from multiple perspectives --- some of which involve bosonic and some fermionic fields.  The question of which is best boils down to which gives a more direct and effective description of the physically important degrees of freedom.

We have focused here on an analogy with the proximity effect in the theory of superconductivity.  The saddle point analysis we have presented suggests (but certainly does not prove) that the QH condensate can extend deep into an otherwise metallic region.  
One might as well attempt a fermionic description of this effect;  in Fig.~\ref{fig: noninteracting} (App.~\ref{app: sec: noninteracting}) we show the local density of states (LDOS) for non-interacting electrons in the presence of a spatially varying magnetic field that interpolates between a value corresponding to the $\nu=1$ QH effect to a zero magnetic field.  By eye, one can see that a peak in the LDOS that one can associate locally with the lowest LL remains well-articulated relatively far into the metallic regime.  This is suggestive of the same basic physics we have found from an alternative perspective, but it is not obvious where to go from this observation.  We thus suggest that the observation of the persistence of QH coherence through a metallic region would constitute a compelling confirmation of the existence in a direct physical sense of a QH condensate.  

Indeed, we were initially inspired to undertake this study by preliminary studies of the Hall response of a 2DEG confined to the surface of a cylinder and subjected to a magnetic field. The field can be arranged so that there are regions of QH fluids separated by metallic regions, which is precisely the geometry we have considered in Sec.~\ref{sec: interface}. For fields such that there is a region of QH fluid on the top and bottom, this experimental geometry is analogous to that of an SNS junction familiar from the study of superconductivity.  It is an ideal geometry for exploring the penetration of a QH condensate into a metallic region.

It is our belief that the two-component CS field theory we developed to treat this problem can be of broader utility.  In this context, we have shown that it permits a physically plausible treatment of the quasielectron and that it may allow a new approach to characterizing charge-neutral collective excitations from the CB perspective.

{\bf Acknowledgement. } We thank Claudia Felser for inspiring discussions that motivated this project.  Work at Stanford (SAK, ZYH, and KSK) was supported in part by the Department of Energy, Office of Basic Energy Sciences, under contract DE-AC02-76SF00515.  SAK was further supported by a Leverhulme Trust International Professorship grant number LIP-202-014 at Oxford.

\bibliography{refs}
\bibliographystyle{apsrev4-1}

\onecolumngrid

\appendix

\section{CB -- CF transmutation} \label{CBCF}

To describe a current with a component perpendicular to the interface, there must be processes that convert CBs into CFs and vice versa. Such processes are also important if we want to describe how the system responds to a magnetic field profile that changes in time, and even for fixed profiles, the relative numbers of CBs and CFs will fluctuate. To describe the CB -- CF transmutation we first recall how the Lagrangian \eqref{eq: Lagrangian} is derived from a first quantized description. 

To go from fermions to the mixed representation we divide the particles in two groups $\{\vec r_i\, ; \ i = 1\dots N_\phi \}$ and  $\{ \vec s_i\, ; \ i = 1\dots N_\psi \}$ and consider fully antisymmetric wave functions
$
\Psi(\vec r_1 \dots \vec s_{N_\psi}) .
$
To illustrate the formalism, we take $(q,q',q'')=(1,0,1)$ which is easy to generalize to arbitrary $(q,q',q'')$ with $q$ odd and $q'$ even. We define a unitary transformation 
\be{unitrans}
U = \exp  i  \left[ \sum_{i < j}^{N_\phi}  \operatorname{arg} (\vec r_i - \vec r_j) + \sum_{i }^{N_\phi} \sum_{j }^{N_\psi}  \operatorname{arg} (\vec r_i - \vec s_j) \right]
\ee
and note
\be{rtrans}
U^\dagger (-i\vec\nabla_{\vec r_i}   +\vec A  )U = (-i\vec\nabla_{\vec r_i}    + \vec a(\vec r_i) +\vec A  )
\ee
where
\be{rgauge}
\vec a (\vec r) = \sum^{N_\phi}_{\vec r_j\ne \vec r}  \vec\nabla_{\vec r}  \operatorname{arg} (\vec r - \vec r_j) + 
 \sum^{N_\psi}_{j}  \vec\nabla_{\vec r}  \operatorname{arg} (\vec r - \vec s_j) 
\ee
and similarly 
\be{strans}
U^\dagger (-i\vec\nabla_{\vec s_i}  +\vec A  )U = (-i\vec\nabla_{\vec s_i}    + \vec b(\vec s_i) + \vec A  )
\ee
where
\be{sgauge}
\vec b (\vec s) = \sum^{N_\phi}_{j}  \vec\nabla_{\vec s}  \operatorname{arg} (\vec s - \vec r_j) \, .
\ee
The corresponding statistical magnetic fields are
\be{bafield}
B_\phi (\vec r) & = & \epsilon^{ij} \partial_i a_j = 2 \pi \sum^{N_\phi}_{\vec r_j\ne \vec r} \delta^2 (\vec r - \vec r_j) 
 + 2 \pi \sum^{N_\psi}_{j} \delta^2 (\vec r - \vec s_j)  \nonumber \\
 & = &   2\pi \rho_\phi(\vec r) + 2\pi \rho_\psi(\vec r)  \\
B_\psi(\vec s) &=& \epsilon^{ij} \partial_i b_j = 2 \pi \sum^{N_\phi}_{j} \delta^2 (\vec s - \vec r_j)
 = 2\pi \rho_\phi(\vec s)  \, . \nonumber
\ee
where the density is a sum of delta functions at the particle positions. These are precisely the constraints in \eqref{fieldcharge} obtained by varying the multiplier fields $a_0$ and $b_0$. Note that the statistical vector potentials $\vec a$ and $\vec b$ have support in the full space. 

So far everything is standard, but we now generalize the transformation \eqref{unitrans} to be time-dependent in such a way that particle $k$ goes from being a composite boson for $t<\tau_k$ to being a fermion for $t>\tau_k$:

\be{tdepunitrans} 
U = \exp i \left[ \sum_{i < j}^{N_\phi}  \operatorname{arg} (\vec r_i - \vec r_j) + \sum_{i\ne k }^{N_\phi} \sum_{j }^{N_\psi}  \operatorname{arg} (\vec r_i - \vec s_j) +   \sum_{j }^{N_\psi}  \operatorname{arg} (\vec r_k - \vec s_j) \theta(\tau_k-t)\right] \, .
\ee
The corresponding statistical $B_\phi$ field is still given by \eqref{bafield}, except that there is no delta function at position $r_k$, while the $B_\psi$ field is,
\be{bbfield}
B_\psi(\vec s) &=& \epsilon^{ij} \partial_i b_j = 2 \pi \sum^{N_\phi}_{i\ne k} \delta^2 (\vec s_i - \vec r_j) + 2 \pi\,  \delta^2 (\vec s_i - \vec r_k) 
\theta(\tau_k-t)\, . \nonumber
\ee
So for $t<\tau_k$ we have the old relations \eqref{bafield}, while for $t>\tau_k$ the CB at $\vec r_k$ has lost its attached unit  of $b$-flux. The statistics are now no longer correct, but this can be fixed if we relabel: $\vec r_k \rightarrow \vec s_{N_\psi + 1}$ so $N_\psi \rightarrow N_\psi +1$. 
Since the newly born CF is already attached to an $a$-flux, the net effect of the transmutation is that the $k^{th}$ CB loses its $b$-flux and is transformed into a CF.

This process is also accompanied by a $\delta$-function pulse of electric circulation.
To see this note that there will be an extra potential term in the Schr\"odinger equation 
\be{insta0}
\tilde b_0  \equiv U^\dagger (t) i\partial_t U(t) = -\sum_{j }^{N_\psi}  \operatorname{arg} (\vec r_k - \vec s_j) \delta(t - \tau_k) 
\ee 
and since we use a static gauge where $\vec b(\vec r)$ is time-independent, the electric field comes entirely from $\tilde b_0$, and all the CFs see a pulse of the electric field,
\be{elpulse}
 \tilde E_\psi(s_j) =  - \hat z \times \frac {\vec s_j - \vec r_k}  {|\vec s_j - \vec r_k|^2} \delta(t - \tau_k) 
 \ee
which is precisely the amount to reflect the instantaneous disappearance of one unit of flux at position a $\vec r_k$ as seen by the particles at positions $\vec s_j$. To see this note
\be{fluxchange}
\Delta \Phi & = & \int_{\tau_k - \epsilon} ^{\tau_k -\epsilon}  \dot \Phi =  \int_{\tau_k - \epsilon}^{\tau_k+ \epsilon} \delta(t - \tau_k)
\oint_{{\mathcal C}_k} d\vec s \cdot \hat z \times  \frac {\vec s - \vec r_k}  {|\vec s- \vec r_k|^2} \delta(t - \tau_k)  \nonumber \, ,  \\
& = &  \int_{{\mathcal C}_k} d\theta =  2\pi \, ,
\ee
where we used Faraday's law in the second equality, and the contour ${\mathcal C}_k$ encircles $\vec r_k$ and is parametrized by the polar angle $\theta$; $\hat z$ is the unit vector perpendicular to the plane. For $q \ne q''$ there would also be a similar instantaneous pulse of electric $E_\phi$ field seen by the remaining CBs. In either case, the correct commutation relation is retained by transforming the CB to a CF as already pointed out.

\section{Detailed discussions on the composite quasiparticles}\label{app: sec: excitations}
\subsection{Formal consideration}\label{app: sec: formal}

To derive the topological properties of the possible composite soliton solutions, we reparameterize the current of CB, $J^\mu_\phi =\frac{1}{2\pi} \epsilon^{\mu\nu\eta}\partial_\nu h_{\eta}$, with a hydrodynamic field $h$ which further couples to the current of the vortices $J_{v}$. In this dual representation, the terms that are relevant to the topological properties read
\begin{align}
    {\cal L}_\text{topo} = \frac{1}{2\pi}(a+A+\frac{q}{2}\omega) \d h + h \cdot J_v + (b+A+\frac{q'}{2}\omega) \cdot J_\psi +{\cal L}_\text{CS}
\end{align}
where we used the short-hand notation $a\cdot b = a_\mu b^\mu$, and introduced the background spin connection of the base manifold, $\omega$, with the prescription introduced in~\cite{PhysRevB.90.115139}. Integrating out $a$ and defining $\beta \equiv  (b- q'' h)$, it takes the form of the Wen-Zee theory~\cite{PhysRevLett.69.953}:
\begin{align}
    {\cal L}_\text{topo} = &\frac{q}{4\pi}  h \d h + \frac{1}{2\pi} h \d\left(\frac{q}{2}\omega+A\right)+h \cdot (J_v+q'' J_\psi)  \nonumber\\
    & \   -\frac{1}{4\pi q'} \beta \d \beta + \left(  \beta +\frac{q'}{2}\omega+A\right)\cdot J_\psi
\end{align}
It thus becomes clear that now we have two decoupled CS fields, and $\beta$ simply attaches $q'$ flux to each CF.

Integrating out $h$ and $\beta$, we get the response action:
\begin{align}
{\cal L}_\text{topo} = &- \frac{1}{4 \pi q} \left(\frac{q}{2}\omega +A\right) \d \left(\frac{q}{2}\omega +A\right) \nonumber\\
    &  - \pi \left(J_v,J_\psi\right)\begin{bmatrix}
       \frac{1}{q}  & \frac{q''}{q} \\
        \frac{q''}{q} & \frac{q''^2-q q'}{q}
    \end{bmatrix} \frac{1}{\d} \begin{pmatrix}
    J_v \\ J_\psi
    \end{pmatrix}\nonumber \\
   & + A \cdot \left[-\frac{1}{q}J_v+\frac{q-q''}{q}J_\psi\right]\nonumber\\
    & + \omega\cdot \left[-\frac{1}{2}J_v+\frac{q'-q''}{2}J_\psi\right]
\end{align}
where the symbol $\frac{1}{\d}$ represents the inverse operator of exterior derivative.

Now let's assume we have found a bound state consisting of $n_v$ vortices and $n_\psi$ CFs, then we may rewrite the response action for the current of these quasi-particles, $J_\text{QP}$, with the substitution $J_v = n_v J_\text{QP}$ and $J_\psi = n_\psi J_\text{QP}$:
\begin{align}
    {\cal L} = & - \frac{1}{4 \pi q} \left(\frac{q}{2}\omega +A\right) \d \left(\frac{q}{2}\omega +A\right) \nonumber\\
    &  + \left[ n_\psi - \frac{n_v+q'' n_\psi}{q} \right]A \cdot J_\text{QP}\nonumber\\
    & - \pi \left[\frac{(n_v+q'' n_\psi)^2}{q} - q' n_\psi^2 \right] J_\text{QP} \frac{1}{\d} J_\text{QP} \nonumber\\
    & + \left[L_\psi^z +  \frac{q' n_\psi - (n_v + q'' n_\psi) }{2} \right] \omega\cdot J_\text{QP}
\end{align}
Note that, besides the angular momentum carried by the gauge fields, we also have included the possible angular momentum $L_\psi^z$ carried by the CF states in such an excitation. Finally, the charge, statistical phase, and angular momentum of this excitation can be read off from the effective action, which are respectively 
\begin{align}
    Q &= - n_\psi + \frac{n_v+q''n_\psi}{q} \\
    \theta_\text{stat} &= q \pi Q^2  \mod 2\pi \\
    S^z &= L_\psi^z +  \frac{q' n_\psi - (n_v + q'' n_\psi) }{2} = L_\psi^z +  \frac{(q'-q) n_\psi -qQ }{2}
\end{align}

\subsection{The saddle point equations} \label{app: sec: saddlepoint}

To obtain the composite quasiparticles, we derive the saddle point equations for $\phi$ and $\psi$:
\begin{align}
    \frac{\delta \mathcal{L}}{\delta \phi^\star } = 0 \implies & \frac{1}{2m} (-\mathrm{i} \nabla + \vec{a}+\vec{A})^2 \phi + V (\rho_\phi+\rho_\psi-\frac{\bar{B}}{2\pi q}) \phi - a_0 \phi = 0 \\
    \frac{\delta \mathcal{L}}{\delta \psi^\dagger } = E_n \psi \implies & \frac{1}{2m} (-\mathrm{i} \nabla + \vec{b}+\vec{A})^2 \psi_n + V (\rho_\phi-\frac{\bar{B}}{2\pi q}) \psi_n - b_0 \psi_n = E_n\psi_n
\end{align}
where all $E_n<0$ states are occupied, and the currents are
\begin{align}
    J^i_{\phi} & = -J_{\phi,i} = \frac{1}{m}\Re \left[\phi^\star (-\mathrm{i}\nabla +\vec{a}+\vec{A})_i \phi  \right]\\
   J^i_{\psi} & = -J_{\psi,i} = \sum_n \frac{1}{m} \Re \left[\psi_n^\star (-\mathrm{i}\nabla +\vec{b}+\vec{A})_i \psi_n \right]
\end{align}
(remember that $A^i = (\vec{A})_i$ in our convention).

In polar coordinates $(r,\theta)$ of the space, we seek rotationally symmetric solutions by decomposing the $\phi$ and $\psi$ field into an amplitude part $f(r)$ and a phase part $\mathrm{e}^{-\mathrm{i}m^z \theta}$. In this coordinate, the uniform background magnetic field can be characterized by $A_\theta \equiv \vec{A} \cdot \hat{e}_\theta = \frac{\bar{B}r}{2} $ (symmetric gauge), and the above equations and the flux attachment conditions translate into 
\begin{align}
\begin{pmatrix}
a_\theta(r) \\ b_\theta(r)
\end{pmatrix} &= - \frac{2\pi}{r} \int_{0}^r d r' ~ r' \mathcal{K} \begin{pmatrix}
f^2_\phi(r') \\  \sum_n f^2_{\psi_n}(r')
\end{pmatrix}\\
\begin{pmatrix}
a_0(r) \\ b_0(r) 
\end{pmatrix} & = \int_r^\infty d r'~ \frac{2\pi }{m}  \mathcal{K} \begin{pmatrix}
f^2_\phi  \left[-\frac{m^z_\phi}{r} + a_\theta(r') + A_\theta(r') \right] \\  \sum_n  f^2_{\psi_n}  \left[-\frac{m^z_{\psi_n}}{r'} + b_\theta(r') + A_\theta(r')\right]
\end{pmatrix}    \\
0 &= \frac{-1}{2m} \left( f''_\phi +\frac{f'_\phi}{r}\right) + \left[\frac{1}{2m} \left(-\frac{m^z_\phi}{r} + a_\theta +A_\theta \right)^2  + V \left(f^2_\phi + \sum_n f^2_\psi - \frac{\bar{B}}{2\pi q} \right) -a_0 \right] f_\phi \\
E_n f_{\psi_n} &= \frac{-1}{2m} \left( f''_{\psi_n} +\frac{f'_{\psi_n}}{r} \right) +\left[\frac{1}{2m} \left(-\frac{m^z_{\psi_n}}{r} +b_\theta +A_\theta\right)^2 + V \left(f^2_\phi - \frac{\bar{B}}{2\pi q} \right)-b_0 \right] f_{\psi_n}
\end{align}
For each set of  $(n_v,n_\psi,L^z_\psi)$ and thus $(q,q',q'')$ we choose, we constrain the value of $m^z_\phi = n_v$ and $\sum_{n=1}^{n_\psi} m^z_{\psi_n} = L^z_\psi$. Then, we solve the coupled differential equations by a mixed-iteration method for both $f_\phi$ ad $f_{\psi_n}$. In each iteration, we first solve $f_\phi$ with a relaxation method, and then solve $f_{\psi_n}$ by diagonalizing the linear operator (which is, eventually, the Hamiltonian for $\psi$). We determine the stability of the solution by inspecting whether the $\psi$ modes we are keeping have negative energy, and they are the only ones that do so, \ie the vortices do trap exactly $n_\psi$ CFs with angular momentum $L^z_\psi$.

After obtaining the solution, the total energy of a soliton can be calculated as 
\begin{align}
    E & = \int d^2 r \frac{1}{2m} \left| \left( -i\vec \nabla+ \vec a  + \vec A \right)\phi\right|^2 + \frac{1}{2m}\left| \left( -i\vec \nabla+ \vec a  + \vec A \right)\psi\right|^2 + V[|\psi|^2,|\phi|^2]\\
    &= \frac{\pi}{m} \int_0^\infty dr ~ r \left\{\left[\left(\frac{d f_\phi}{ d r}\right)^2 + f_\phi^2 \left(-\frac{m^z_\phi }{r} + a_\theta+A_\theta\right)^2\right] \right. \nonumber\\
    & \ \ \ \  \ \ \ \ \ \ \ \ \ \ \ \ \ \ \ \ \ \ \ \ + \left. \sum_n \left[\left(\frac{d f_{\psi_n}}{ d r}\right)^2 + f_{\psi_n}^2 \left(-\frac{m^z_{\psi_n} }{r}+ b_\theta + A_\theta\right)^2\right]+ 2m V[\rho_\phi, \rho_\psi]\right\}
\end{align}

\subsection{Relation to the CFT quasielectron} \label{app: sec: CFT}

\newcommand\hole[1] {H_{\frac{1}{q}}(\eta_#1)}
\newcommand\vme {e^{i\sqrt{q} \varphi (z)}} 
\newcommand\vm[1] {V_1(z_{#1}) }
\newcommand\nvm[1] {V(z_{#1}) }
\newcommand\vtme {\partial e^{i(\sqrt{q}-\frac 1 {\sqrt q}) \varphi (z)}}

In the CFT approach, QH wave functions are expressed as correlators of operators that represent electrons and quasiparticles~\cite{moore1991nonabelions}. In the simplest case, these operators are vertex operators in a scalar CFT, but in other cases, they can be combined with Majorana or parafermion operators. With this in mind, we define the normal-ordered vertex operators,
\be{vo}
V(z) &=&\, : \vme : \\
H_{1/q}(\eta) &=&\, :e^{ \frac i {\sqrt q} \varphi (\eta) } : \label{hole} \, ,
\ee
where the normal ordering symbol $:\ \ :$, will be suppressed in the following. 
The free massless boson field,    $\varphi$, is    normalized so as to 
have the (holomorphic) two-point function
\be{twop}
\av{ \varphi (z) \varphi (w) } = - \ln (z - w)   \, ,
\ee
so that the vertex operators obey the relations 
\be{verrel}
e^{i\alpha \varphi(z)}  e^{i\beta \varphi(w)}
	 &=& e^{i\pi \alpha\beta}e^{i\beta \varphi(w)} e^{i\alpha \varphi(z)} 
	= (z-w)^{\alpha\beta} e^{i\alpha \varphi(z) + i\beta \varphi(w)} \nonumber \\
	&\sim &(z-w)^{\alpha\beta} e^{i(\alpha  + \beta) \varphi(w)} 
\ee	 
With this, the (holomorphic) wave function for a $N$ electrons and $n$ quasiholes can be expressed as a CFT correlator 
\be{lqh}
\Psi_L (\eta_1\dots \eta_n ; z_i) =  \av{ \hole 1 \hole 2 \dots \hole n \nvm 1  \nvm 2 \dots \nvm {N-1} \nvm N } .
\ee

A natural guess for a quasielectron operator is to just change the sign in the exponent in the quasihole operator \ie to use $e^{-\frac i {\sqrt q} \varphi(\eta)}$.  However, as discussed in the text, such an anti-vortex introduces unacceptable singular terms $\sim \prod_i (z_i - \eta)^{-1}$ in the electronic wave function. Inspired by the CF wave functions, we instead define a quasielectron operator, $ P_{\frac{1}{q}}(z)$, that will {\em replace} one of the original electron operators $V(z)$. Thus, $P(z)$ is a modified electron operator, with a different amount of vorticity. 
 The excess electric charge associated with such a modification is the difference between the charges of the operators $V$ and $P_{\frac{1}{q}}$ \ie  $\Delta Q_{el} =  e((1-1/q)-1) = -e/q$, as appropriate for a quasielectron
at $\nu=1/q$. The modified electron  operator is taken to be
\be{qpo}
 P_{1/q} (z) = \vtme ,
\ee
where the derivative is put in by hand in order for the correlators of $P_{1/q}(z)$ with a number of $V(z_i)$ not to vanish under antisymmetrization. Clearly one would like to eliminate this \emph{ad hoc} assumption. 

Let us now specialize to $q=3$, and fuse  $P_{1/3}(z)$ with an electron operator using \eqref{verrel},
\be{twoelfusion}
P_{1/q}(z) V(z_i) \sim \partial_z (z-z_i)^2\, e^{i\frac 5 {\sqrt 3} \varphi(z_i)} \, .
\ee
The interpretation of this formula is again obtained by comparing it with the ground state, where there are two electrons at $z$ and $z_i$ and 6 associated vortices. The fused operator has again two electrons, but only 5 vortices so there is effectively -1 vortex. In addition, there is the factor $(z-z_i)^2$ which in the field theory is interpreted as $q'=2$ and one derivative, which in our calculation amounts to the electron at $z$ being in a p-wave. Thus our fused quasielectron operator has the same $n_v, n_\psi$ and $L^Z$ as the favored $-1,2, 1$ solution in the text. 

\section{Details on the QH-Metal interface problem}\label{app: sec: interface}

We start our analysis by adopting the dual representation of $\phi$ theory. To do so, we decompose $\phi = \sqrt{\rho_\phi}\mathrm{e}^{-\mathrm{i}\varphi}$, and introduce a Hubbard-Stratanovich field $\vec{j}$ to rewrite:
\begin{align}\label{app: hydro}
   \mathcal{L}_{\phi}[\phi^\star,\phi;a] & \rightarrow \mathcal{L}_{\phi}[\rho_\phi,\varphi;a] = \rho_\phi \left[-\partial_0 \varphi +a_0+ A_0\right] -\frac{\rho_\phi}{2m}\left(-\nabla \varphi +\vec{a} +\vec{A}\right)^2 - \frac{(\nabla \rho_\phi)^2}{8m\rho_\phi} \nonumber\\
   & \rightarrow \mathcal{L}_{\phi}[\rho_\phi,\vec{j},\varphi;a] = \rho_\phi \left[-\partial_0  \varphi+a_0+A_0\right] -\vec{j}\cdot \left(-\nabla \varphi +\vec{a} +\vec{A}\right) + \frac{m\vec{j}^2}{2\rho_\phi} - \frac{(\nabla \rho_\phi)^2}{8m\rho_\phi}
\end{align}
We thus see that $(\vec{j})_i$ is nothing but the CB current $J^i_\phi = \frac{\rho_\phi}{m} \left(-\nabla \varphi+\vec{a}+\vec{A}\right)_i$. Integrating out $\varphi$ yields the continuity equation $\partial_\mu J^\mu_\phi = 0$, and allows us to further parametrize $J_\phi^\mu =\epsilon^{\mu\nu\sigma} \partial_\nu h_\sigma/(2\pi)$ with a hydrodynamic field $h$:
\begin{align}
   \mathcal{L}_{\phi}[h,a] & = \frac{1}{2\pi}(a+A)\d h - \frac{m \vec{E}_h^2}{4\pi B_h} +\frac{(\nabla B_h)^2}{16\pi mB_h}
\end{align}
where we use $B_h \equiv - (\partial_1 h_2-\partial_2 h_1) = -2\pi \rho_\phi$ and $(\vec{E}_{h})_i \equiv - (\partial_i h_0-\partial_0 h_i) =2\pi  \epsilon_{ij}J^j_\phi  $ to represent the effective magnetic and electric fields if viewing $h$ as a gauge field. We note that by rewriting the theory into this form, we have assumed that the CBs remain condensed and vortices are expelled,  which is justifiable so long as the effective magnetic field seen by CBs is small compared to the superfluid density.

Then, with this dual representation, we put everything together, integrate out $a$ and shift $b\rightarrow b+q'' h$. Finally, we get:
\begin{align}
    \mathcal{L}[\psi^\dagger,\psi;h,b] =& \mathcal{L}_\text{CF}[\psi^\dagger,\psi;q''h+b]   -\frac{1}{4\pi q'} b \d b  \nonumber\\
    & \ \ \ -\frac{V_1B_h^2}{4\pi^2} + V_2 \frac{B_h}{2\pi} |\psi|^2 + A_0 \left( |\psi|^2 - \frac{B_h}{2\pi}\right) -\frac{m\vec{E}^2_h}{4\pi B_h} + \frac{(\nabla B_h)^2}{16\pi m B_h} +\frac{q}{4\pi} h\d h + \frac{1}{2\pi} A \d h \label{app: dualLagrangian}
\end{align}
We see that $b$ still attaches $q'$ flux to each CF. Below we will seek steady-state solutions for spatially non-uniform magnetic fields $B(\vec{r})$ with no electric field, such that all the time dependence in the Lagrangian can be dropped.

We note that the treatment of the CFs is very difficult so the best we could do is to treat $\mathcal{L}_\text{CF}$ effectively. Specifically, we will make the following key assumptions to simplify the theory: 
\begin{itemize}
    \item  The CFs form a dissipative fermi liquid that cannot host any persisting current,
\begin{align}
    J_\psi^i =0, \label{app: nocurrent}
\end{align}
when considering phenomena at length scales above the mean free path of the CFs, $\ell_\text{mf}$. 
\item The long-range Coulomb interaction is effectively keeping the overall electron density remains fixed to be 
\begin{align} \label{app: constantrho}
    \rho_\phi+\rho_\psi = \bar{\rho}, 
\end{align}
when considering phenomena at length scales above a characteristic one, $\ell_{\rho}$.
\item The kinetic energy density of the CFs is equal to that of a Fermi liquid with the same density, \ie
\begin{align}
\mathcal{E}_\text{kin} = \rho_\psi^2/2\kappa
\end{align}
where $\kappa$ is the compressibility of the CFs. As an estimate, for non-interacting fermion gas, $\kappa = \frac{m}{2\pi}$. We note that this term takes the form of a potential, and effectively provides a ``Fermi degenerate pressure''.
\end{itemize}

The constant density constraint, Eqs.~\ref{app: constantrho}, can be imposed by a lagrangian multiplier $\eta$. Then, with those assumptions, we write down an effective Lagrangian justifiable at a large length scale:
\begin{align}
    \mathcal{L}_\text{eff}[J_\psi; h,b;\eta] =& J_\psi \cdot (q'' h + b + A) + \eta(\rho_\phi+\rho_\psi-\bar{\rho}) 
     -\frac{1}{4\pi q'} b \d b   - \frac{\rho_\psi^2}{2\kappa }  \nonumber\\
    & \ \ \ -V_1 \rho_\phi^2/2 -V_2 \rho_\phi\rho_\psi + \frac{m(\nabla h_0)^2}{2 (2\pi)^2 \rho_\phi} - \frac{(\nabla \rho_\phi)^2}{8 m \rho_\phi} +\frac{q}{4\pi} h\d h + \frac{1}{2\pi} A \d h \label{app: effectiveLagrangian}
\end{align}
where $\rho_\phi = -\frac{B_h}{2\pi} \equiv \frac{1}{2\pi} (\partial_1 h_2-\partial_2 h_1)$ should be understood. We further note that in the absence of a background electric field, $A_0$ is a constant and can be absorbed into $\eta$.

Starting from this effective Lagrangian, we analyze the saddle point equations of $h, J_\psi$ under the ``no $\phi$ current'' and ``constant density'' conditions in Eqs.~\ref{app: nocurrent}\&\ref{app: constantrho}:
\begin{align} 
    \frac{\delta \mathcal{L}}{\delta h_0} =0 
     &\implies q'' \rho_\psi + q\rho_\phi - \frac{B}{2\pi} - \frac{m}{(2\pi)^2} \nabla(\frac{1}{\rho_\phi}\nabla h_0) = 0\label{app: h0EOM}\\
    \frac{\delta \mathcal{L}}{\delta h_i} =0 &\implies  \epsilon^{ij} \partial_j  \left\{ qh_0 + \eta - V_1 \rho_\phi -V_2\rho_\psi - \frac{m^2(\nabla h_0)^2}{2(2\pi)^2 \rho_\phi^2}+\frac{1}{8 m } \left[ 2\nabla( \frac{1}{\rho_\phi}\nabla \rho_\phi) + \frac{(\nabla \rho_\phi)^2}{\rho_\phi^2} \right]  \right\}  = 0 \label{app: hiEOM} \\
     \frac{\delta \mathcal{L}}{\delta \rho_\psi} =0 
     &\implies q'' h_0 + b_0 +\eta -\rho_\psi/\kappa -V_2 \rho_\phi =0 \label{app: rhopsiEOM}\\
      \frac{\delta \mathcal{L}}{\delta J^i_\psi} =0 
     &\implies q'' h_i + b_i +A_i  = 0  \label{app: JpsiEOM} \\
     \frac{\delta \mathcal{L}}{\delta b_0} =0 
     &\implies \rho_\psi - \frac{1}{2\pi q'} \epsilon^{ij}\partial_i b_j =0 \label{app: b0EOM}\\
      \frac{\delta \mathcal{L}}{\delta b_i} =0 
     &\implies - \frac{1}{2\pi q'} \epsilon^{ij}\partial_j b_0 =0 \label{app: biEOM} 
\end{align}

From Eq.~\ref{app: rhopsiEOM} and \ref{app: hiEOM}, one can see that $\eta_0$ has the form of an electric potential seen by both $\phi$ and $\psi$; in fact, at saddle-point level, it can be viewed as the potential provided by the long-range interaction
\begin{align}
    \eta_0(\vec{r}) = - \frac{\delta V_\text{long range}[\rho] }{\delta \rho(\vec{r})}
\end{align}
which is presumably too complicated to be treated in an {\it ab initio} way such that we could only impose its consequences effectively. With this understanding, we see that the CFs see a constant overall potential, implying that no force is applied to the $\psi$ and verifying that there is no $\psi$ current due to the drifting motions.

Combining Eqs.~\ref{app: h0EOM}, \ref{app: hiEOM}, \ref{app: rhopsiEOM}\&\ref{app: biEOM}, one can obtain the below differential equations for the two independent unknowns $\rho_\phi(\vec{r})$ (which derives $\rho_\psi = \bar{\rho} - \rho_\phi$) and $h_0(\vec{r})$ (which derives $J^i_\phi(\vec{r}) = \epsilon^{ij} \frac{\partial_j h_0}{2\pi }$), for any $B(\vec{r})$ configuration:
\begin{align}
    &\frac{m}{(2\pi)^2} \nabla\left(\frac{1}{\rho_\phi}\nabla h_0 \right) = q''\bar{\rho} + (q-q'')\rho_\phi - \frac{B}{2\pi}  \label{app: differentialequation1}  \\
    &\frac{1}{4\sqrt{\rho_\phi}}  \nabla\left(\frac{1}{\sqrt{\rho_\phi}}\nabla \rho_\phi \right) =  m\tilde{V}(\bar{\rho}-\rho_\phi) +(q''-q)mh_0 + \frac{m^2 (\nabla h_0)^2}{2 (2\pi \rho_\phi)^2} \label{app: differentialequation2}
\end{align}
where we have defined $\tilde{V} \equiv 2V_2-V_1-1/\kappa$ and fixed an undetermined constant to make the equations reasonable in a pure CB region with constant $B = 2\pi q \bar{\rho}$ and $h_0=0$.

Alternatively, the problem can be solved by the minimization of energy functional (with all the constraints understood):
\begin{align}
    \mathcal{E}[\rho_\phi, h_0] = \frac{\left(\nabla \rho_\phi\right)^2}{8m\rho_\phi} - \frac{m\left(\nabla h_0\right)^2}{2(2\pi)^2\rho_\phi} + h_0\left[\frac{B}{2\pi} - q''\bar{\rho} - (q-q'')\rho_\phi\right] - \frac{\tilde{V}}{2} \left(\bar{\rho}-\rho_\phi\right)^2
\end{align}

The above results apply to general $B(\vec{r})$. Next we consider an interface problem where $B(x,y)$ is assumed to only depend on $x$ coordinate. We assume $B\rightarrow 2\pi q \bar{\rho}$ at $x\rightarrow -\infty$ and $B\rightarrow 2\pi q' \bar{\rho}$ at $x\rightarrow +\infty$, such that deep in the $x\rightarrow \mp \infty$ regimes, the system can be purely described by a CB condensate and a CF liquid, respectively.

To find general solutions for $\rho_\phi$ profile in this case, we consider the limit where the interface is wide, \ie $B(x)$ varies slowly at a length scale $L$ that is much greater than all the microscopic length scales such as magnetic length $\ell_B$, mean free path of the CFs $\ell_\text{mf}$, and the density fluctuation length $\ell_\rho$. For later convenience, we make the decomposition
\begin{align} \label{app: eq: rhoB definition}
    &\rho_\phi(x) = \rho_B(x) +\delta \rho(x) \\
    \rho_B&\equiv \left(\frac{B}{2\pi} -q''\bar{\rho}\right)/(q-q'') 
\end{align}
The physical meaning of $\rho_B$ is that if $\rho_\phi$ strictly follows $\rho_B$, the effective flux seen by the $\phi$ particles vanishes. 

Integrating Eq.~\ref{app: differentialequation1}, we can obtain the formal solution of $h_0(x)$ as a functional of $\rho_\phi(x)$:
\begin{align}
    h_0(x) =& (q-q'') (2\pi)^2/m \int_{-\infty}^{x} \mathrm{d}x' \rho_\phi(x') \int_{-\infty}^{x''} \mathrm{d} x'' \delta\rho(x'') \\
    =& (q-q'') (2\pi)^2/m \left\{ \int_{-\infty}^{x} \mathrm{d}x' \rho_B(x') \int_{-\infty}^{x''} \mathrm{d} x'' \delta\rho(x'') + \frac{1}{2} \left[\int_{-\infty}^{x} \mathrm{d}x' \delta \rho(x') \right]^2\right\}
\end{align}
Substituting this formal solution into Eq.~\ref{app: differentialequation2}, we obtain an integro-differential equation for $\rho_\phi$:
\begin{align}\label{app: integro differential}
    \frac{1}{4} 
    \left[\frac{\ddot \rho_\phi}{\rho_\phi} - \frac{(\dot\rho_\phi)^2}{2\rho_\phi^2}\right] =  m\tilde{V}(\bar{\rho}-\rho_\phi) - (q-q'')^2(2\pi)^2 \int_{-\infty}^{x} \mathrm{d}x' \rho_B(x') \int_{-\infty}^{x''} \mathrm{d} x'' \delta\rho(x'')
\end{align}
This equation is in general hard to solve, but now we use the following two approximations to simplify it. 
\begin{itemize}
    \item Noting that $L\gg \ell_B \sim 1/\sqrt{\rho}$, we drop the terms on the left hand side, since they are suppressed by $\ell_B^2/L^2$.
    \item We recall that we have made the assumption that there is no vortex in CB condensate, which validates the formalism in Eq.~\ref{app: hydro}. This requires that the magnitude of the effective flux density seen by CB, which is equal to $|(q-q'')\delta \rho|\sim |\delta\rho|$, is much smaller than $\rho_\phi$, so that we can approximate $\rho_\phi$ with $\rho_B$. We note that to give the correct asymptotic behavior $\rho_\phi\rightarrow 0$ as $B\rightarrow 2\pi q' \bar{\rho}$ at $x\rightarrow \infty$, we necessarily need to take $q'' =q'$.
\end{itemize}
 Those approximations, We which we will justify later, reduce the equation to
\begin{align}
   m\tilde{V}(\bar{\rho}-\rho_B) = (q-q'')^2(2\pi)^2 \int_{-\infty}^{x} \mathrm{d}x' \rho_B(x') \int_{-\infty}^{x''} \mathrm{d} x'' \delta\rho(x'')
\end{align}
which is exactly solved by 
\begin{align}
    \delta \rho = - \frac{m\tilde{V}}{(q-q
    '')^2(2\pi)^2} \frac{\mathrm{d}^2\left(\ln \rho_B \right)}{(\mathrm{d}x)^2}
\end{align}
which also leads to
\begin{align}
    &h_0 \approx - \frac{\tilde{V}}{(q-q'')} (\rho_{B}-\bar{\rho}) \\
    \implies &J_\phi^y = -\frac{1}{(2\pi)} \partial_x h_0 =  \frac{\tilde{V} \dot{\rho}_B }{2\pi(q-q'')}  \\
    \implies & I^y = \int_{-\infty}^{\infty} \mathrm{d}x J_\phi^y(x) = 
 - \frac{\tilde{V}\bar{\rho}}{2\pi(q-q'')}
\end{align}

We now more rigorously justify the assumptions in obtaining this solution. In order to safely neglect the left hand side of Eq.~\ref{app: integro differential}, it is necessary that $m\tilde{V}(\Bar{\rho}-\rho_\phi) \gg 1/L^2$. In order to safely approximate $\rho_\phi$ with $\rho_B$, we need $|\delta\rho| \ll \rho_\phi$, which implies $|m\tilde{V}|/L^2 \ll \rho_\phi$. Combining these two conditions, we are able to justify the approximate solution as long as
\begin{align}\label{app: eq: mV conditions}
    \frac{1}{L^2 (\bar{\rho} - \rho_\phi)} \ll |m\tilde{V}| \ll L^2 \rho_\phi
\end{align}
Given $L\gg \ell_B$, this is always satisfied as long as $\rho_\phi,\rho_\psi$ are not too small compared to $\bar{\rho}$.

\begin{figure}
    \centering
    \includegraphics[width = 0.45 \linewidth]{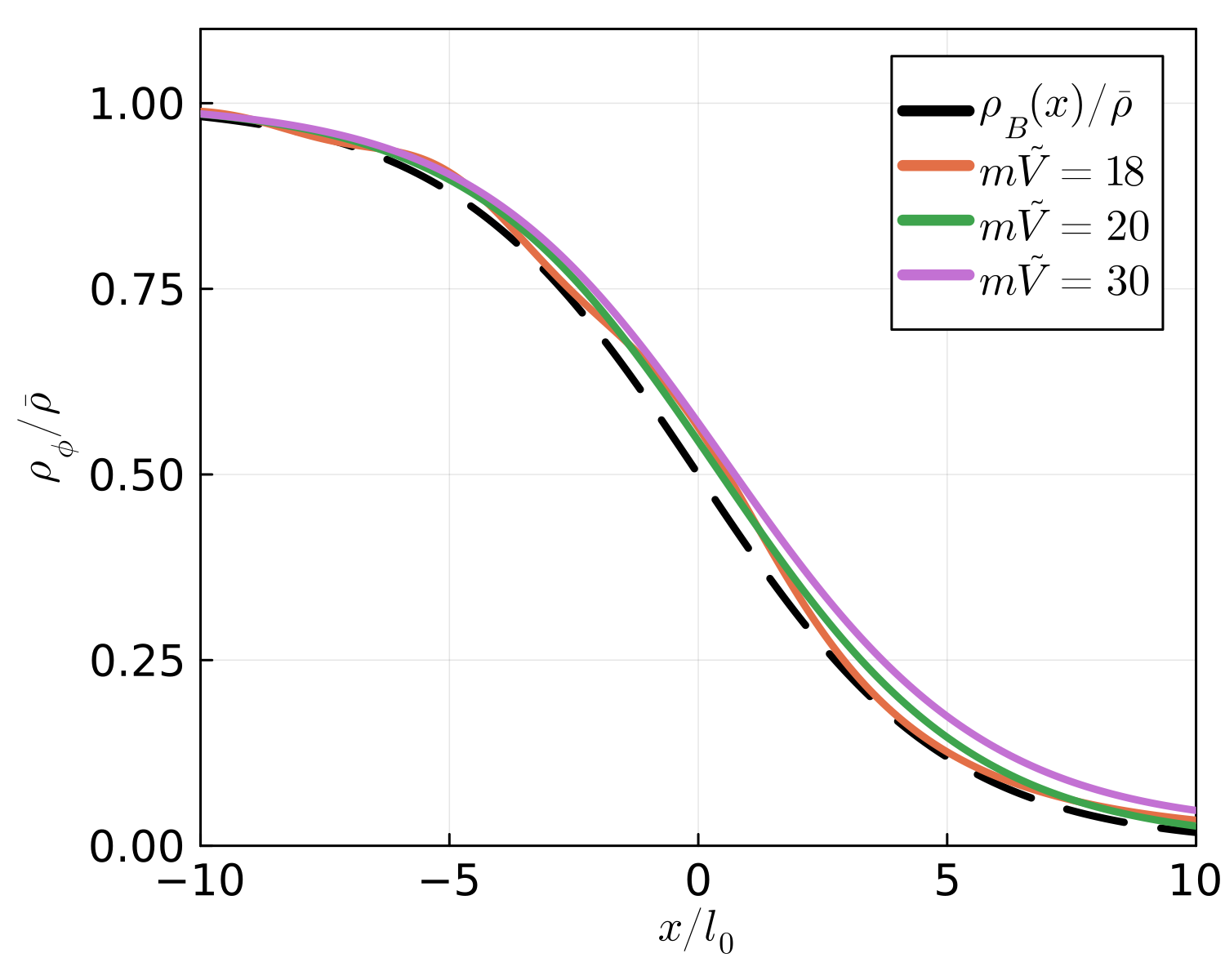}
    \caption{The density profile $\rho_\phi(x)$ solved from Eqs.~\ref{app: differentialequation1} and \ref{app: differentialequation2} with $q=3$, $q'=q''=0$ for magnetic field profile $B(x)= B_0 \left(1-\tanh\frac{x}{5l_0}\right)/2$, where $l_0=1/\sqrt{B_0}$ and $\bar \rho=B_0/(2\pi q )$ are respectively the magnetic length and the electron density deep in the FQH regime $x\rightarrow -\infty$. $\rho_B(x)/\bar{\rho} = B(x)/B_0$ is plotted in black as a reference (see Eq.~\ref{app: eq: rhoB definition} for definition of $\rho_B$). }
    \label{fig: densityprofile_q3}
\end{figure}

To verify the qualitative correctness of the solution in Eq.~\ref{eq: delta rho solution}, 
we numerically analyze Eq.~\ref{eq: integro differential} with $q=3$, $q''=q'=0$. Indeed, for a specific magnetic field configuration with a finite interface width, as illustrated in Fig.~\ref{fig: densityprofile_q3}, we do find that $\rho_\phi(x)$ roughly follow $\rho_B(x)$ when Eq.~\ref{app: eq: mV conditions} is satisfied. 

\begin{figure}
    \centering
   \subfigure[]{\includegraphics[width = 0.49 \linewidth]{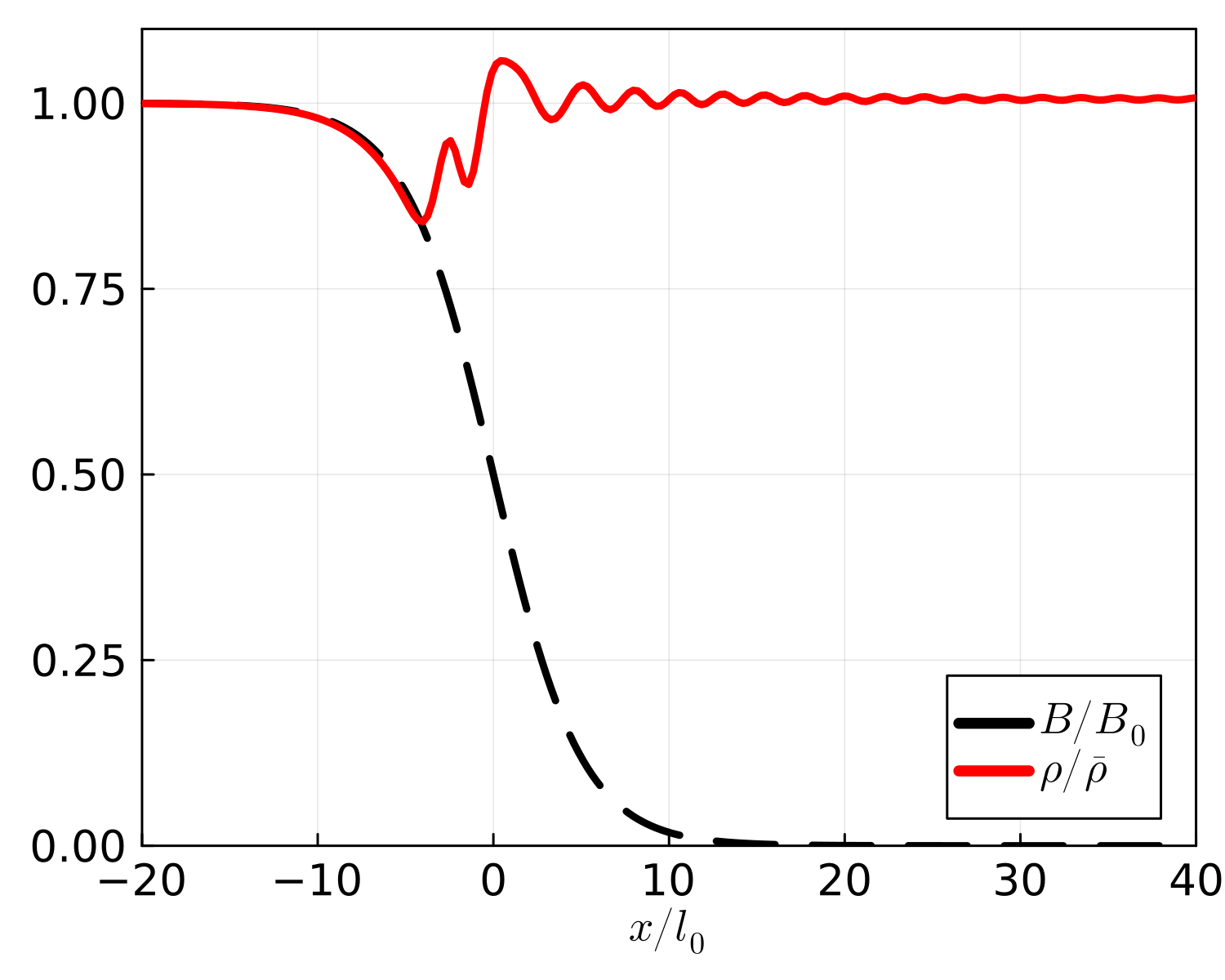}    \label{fig: noninteracting_density_profile}}
    \subfigure[]{\includegraphics[width = 0.49 \linewidth]{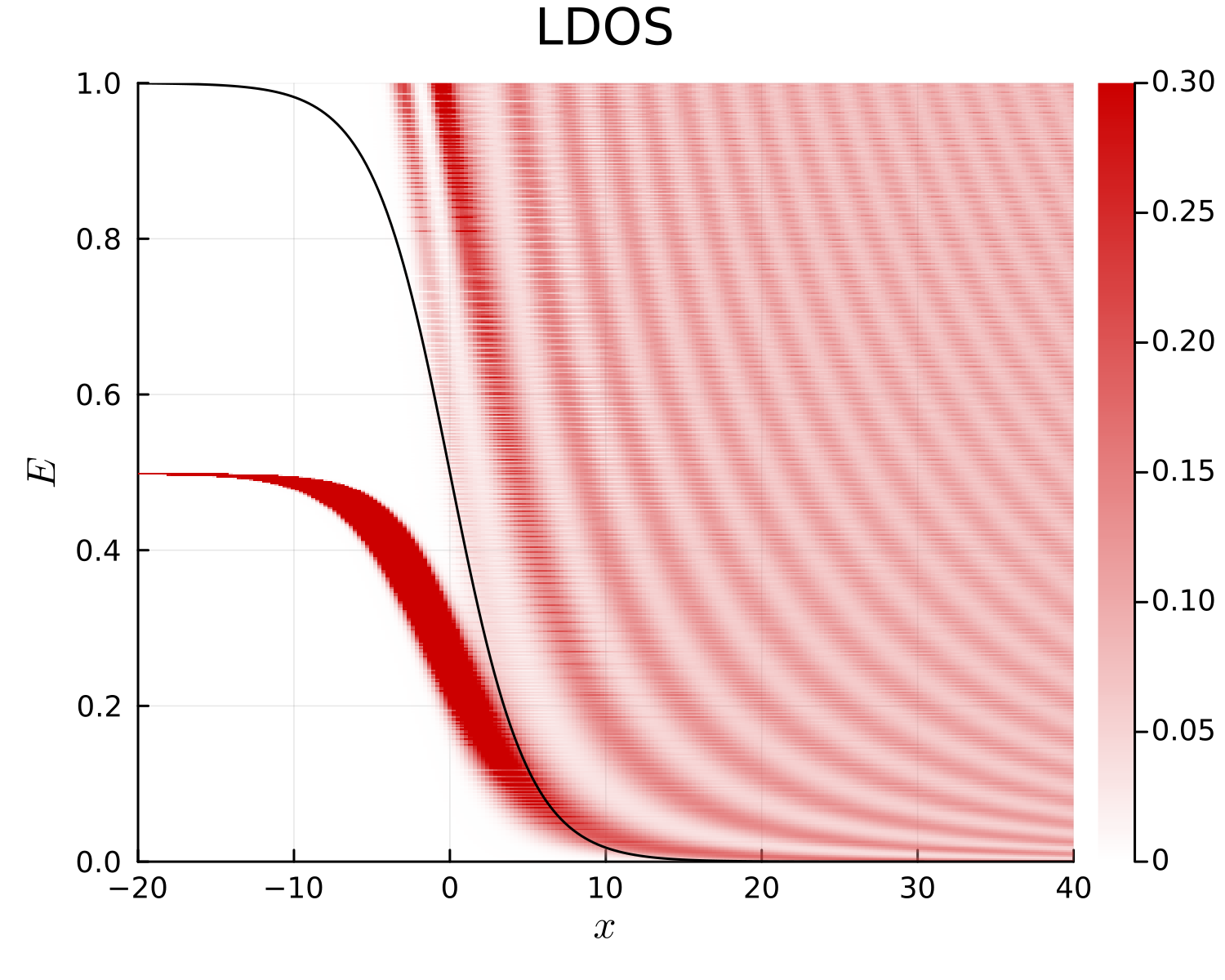}    \label{fig: LDOS} }
    \caption{ Interface between a $\nu=1$ integer QH and a free metal corresponding to the magnetic field profile $B(x)= B_0 \left[1-\tanh(\frac{x}{5l_0})\right]/2$ and chemical potential $\mu =\omega_0$, where $l_0=1/\sqrt{B_0}$ and $\omega_0 =B_0/m$ are respectively the magnetic length and cyclotron energy deep in the QH regime at $x\rightarrow -\infty$. This choice of chemical potential sets the electron density to be $\rho \rightarrow \bar \rho=1/(2\pi l_0^2)$ for both $x\rightarrow \pm \infty$ limits. The plots show (a) the density profile $\rho(x)/\bar \rho$ at fixed $\mu$ with $B(x)/B_0$ in black dashed line shown for comparison;  (b) the local density of states (LDOS), measured in units of $1/(\omega_0l_0^2)$, for energies below $\mu$, with the local cyclotron energy $\omega_c(x)/\omega_0 \equiv B(x)/B_0$ again shown in black for comparison. }
    \label{fig: noninteracting}
\end{figure}

\section{Non-interacting interface}
\label{app: sec: noninteracting}

As a guide to intuition, we have analyzed the structure of the QH to metal interface in the non-interacting limit where the single-particle states can be computed exactly.   
In Fig.~\ref{fig: noninteracting} we show the numerically computed density profile as well as the local density of states (LDOS) of a non-interacting system with a spatially varying magnetic field, which is assumed to have only $x$ dependence in a way that $x\rightarrow -\infty$ region is an integer QH with $\nu=1$ and $x\rightarrow \infty$ region is a free metal. As can be seen in Fig.~\ref{fig: noninteracting_density_profile}, the density at fixed chemical potential (chosen so that the density approaches identical limits as $x \to \pm \infty$) is not strongly modulated across the interface even before adding  any interactions. In Fig.~\ref{fig: LDOS}, we see the lowest LL remains identifiable as a peak in the LDOS far into the region in which there are many higher occupied LLs, which intuitively justifies our  picture of a CB condensate extending into a metallic region.

\end{document}